\documentclass[12pt,a4paper]{article}
\pdfoutput=1
\usepackage[latin1]{inputenc}
\usepackage[T1]{fontenc}
\usepackage{amsfonts,amsbsy,bm,euscript,mathrsfs}
\usepackage{amssymb,faktor,slashed}
\usepackage{color}
\usepackage[tbtags]{amsmath}
\usepackage[bookmarks=true,colorlinks=true,linkcolor=black,citecolor=black,urlcolor=black,bookmarksnumbered,linktocpage=true]{hyperref}
\usepackage{graphicx}

\usepackage[a4paper,text={170mm,257mm},centering]{geometry}

\numberwithin{equation}{section}

\makeatletter
\renewcommand\section{\@startsection {section}{1}{\z@}%
{-3.5ex \@plus -1ex \@minus -.2ex}%
{2.3ex \@plus.2ex}%
{\normalfont\large\bfseries}}
\renewcommand\subsection{\@startsection{subsection}{2}{\z@}%
{-3.25ex\@plus -1ex \@minus -.2ex}%
{1.5ex \@plus .2ex}%
{\normalfont\normalsize\bfseries}}
\makeatother

\expandafter\def\expandafter\bfseries\expandafter{\bfseries\ifmmode\else\boldmath\fi}
\expandafter\def\expandafter\mdseries\expandafter{\mdseries\ifmmode\else\unboldmath\fi}
\expandafter\def\expandafter\normalfont\expandafter{\normalfont\ifmmode\else\unboldmath\fi}

\providecommand{\href}[2]{#2}
\newcommand{\arxivlink}[1]{\href{http://arxiv.org/abs/#1}{[arXiv:#1]}}
\newcommand{\doilink}[2]{\href{http://doi.org/#2}{#1}}

\newcommand{\mathsym}[1]{{}}
\def\id{\protect{{1 \kern-.28em{\rm l}}}}
\def\be{\begin{eqnarray}}
\def\ee{\end{eqnarray}}

\def\tr{{\rm tr}}
\def\ha{\tfrac{1}{2}}
\def\td{\tilde}

\def\S{{\mathcal S} }

\def\z{\zeta}

\def\a{\alpha}
\def\b{\beta}

\def\g{\gamma}

\def\w{\omega}

\def\k{\kappa}

\def\det{\hbox{det}}

\def\Tr{{\rm Tr}}

\def\l {\lambda}
\def\bl{{\tilde \l}}

\def\O{{\mathcal O}}
\def\m{\mu}

\def\foot{\footnote}
\newcommand{\rf}[1]{(\ref{#1})}

\def\F{{\cal F}}

\def\no{\nonumber}

\def\la{\label}
\def\l{\lambda}
\def\tl{{\tilde \l}}
\def\tk{{\tilde k}}

\def\r{\rho}

\def\varpi{{\rm w}}

\def\ve{\varepsilon}

\def\del{\partial}
\def\s{\sigma}
\def\eps{{\epsilon}}
\def\n{\nu}

\def\ed{\end{document}}

\def\iffa{\iffalse}

\def\ep{\epsilon}

\def\ad{{\rm ad}}
\def\d{\delta}

\def\L{\mathcal{L} }

\def\sm{$\sigma$-model}
\def\sms{$\sigma$-models}

\def\bl{\bar \lambda}
\def\Lie{\operatorname{Lie}}

\def\k{\varkappa} 

\def\ddt{\frac{d}{dt}}

\def\kk{{\rm k}}
\def\ka{{\kappa}}

\def\bh{\bar{h}}
\def\lam{$\lambda$-model}

\def\bh{\bar h}
\def\hh{{\rm h}}
\def\PCM{\L_{\rm PCM}}
\def\WZ{\L_{\rm WZ}}
\def\M{{\cal M}}
\def\q{{\rm q}}
\def\RR{{J}}
\def\LL{{K}}

\def\ed{\end{document}}
\def\F{H}
\def\cg{c_{_G}}
\def\cf{c_{_\F}}
\def\cff{c_{_\F}}
\def\HH{\F}
\def\XX{{X}}
\def\hhh{\widetilde {\hh}}
\def\e{\varepsilon}
\def\tih{{\hat h}}
\def\H{{\rm H}}
\def\GG{{\rm G}}
\def\BB{{\rm B}}
\def\gg{\gamma}
\def\bl{\bar\lambda}

\begin{document}


\begin{flushright}\small{Imperial-TP-AT-2019-{06}}

\end{flushright}

\vspace{2.5cm}

\begin{center}

{\Large\bf Integrable sigma models and 2-loop RG flow}

\vspace{1.5cm}

{
Ben Hoare$^{a,}$\footnote{\ bhoare@ethz.ch}, \
Nat Levine$^{b,}$\footnote{\ n.levine17@imperial.ac.uk} and
Arkady A. Tseytlin$^{b,}$\footnote{\ Also at the Institute of Theoretical and Mathematical Physics, MSU and Lebedev Institute, Moscow.
\\\hspace*{15pt} \ tseytlin@imperial.ac.uk}
}

\vspace{0.5cm}

{
\em \vspace{0.15cm}
$^{a}$Institut f\"ur Theoretische Physik, ETH Z\"urich,\\
\vspace{0.05cm}
Wolfgang-Pauli-Strasse 27, 8093 Z\"urich, Switzerland.
\\
\vspace{0.15cm}
$^{b}$Blackett Laboratory, Imperial College, London SW7 2AZ, U.K.
}
\end{center}

\vspace{0.5cm}

\begin{abstract}
Following \href{http://arxiv.org/abs/1907.04737}{arXiv:1907.04737}, we continue our investigation of the relation between the renormalizability (with finitely many couplings) and integrability in 2d $\sigma$-models. We focus on the ``$\lambda$-model,'' an
integrable model associated to a group or symmetric space and containing as special limits a (gauged) WZW model
and an ``interpolating model'' for non-abelian duality. The parameters are the WZ level $k$ and the coupling $\l$, and the fields are $g$, valued in a group $G$, and a 2d vector $A_\pm$ in the corresponding algebra. We formulate the $\lambda$-model as a $\sigma$-model on an extended $G\times G \times G $ configuration space $(g, h, \bar h)$, defining $h $ and $\bar h$ by $A_+ = h \del_+ h^{-1},\ A_- = \bar h \del_- \bar h^{-1}$. Our central observation is that the model on this extended configuration space is renormalizable without any deformation, with only $\lambda$ running. This is in contrast to the standard $\sigma$-model found
by integrating out $A_\pm$, whose 2-loop renormalizability is only obtained after the addition of specific finite local counterterms, resulting in a quantum deformation of the target space geometry. We compute the 2-loop $\beta$-function of the $\lambda$-model for general group
and symmetric spaces, and illustrate our results on the examples of $SU(2)/U(1)$ and $SU(2)$. Similar conclusions apply in the non-abelian dual limit implying that non-abelian duality commutes with the RG flow. We also find the 2-loop $\beta$-function of a ``squashed'' principal chiral model.
\end{abstract}

\newpage

\tableofcontents

\setcounter{footnote}{0}
\setcounter{section}{0}

\section{Introduction}

Certain 2d $\s$-models have the special property of
\textit{renormalizablility}, meaning
they have only finitely many couplings running under RG flow. This property is
expected to be closely connected with integrability \cite{Fateev:1992tk}: the
conservation of infinitely many hidden symmetry charges should reduce the
RG flow in the infinite-dimensional space of \sm{} couplings to a
finite-dimensional one. Having previously been observed
\cite{Fateev:1992tk,Fateev:2019xuq} only at the 1-loop (Ricci flow) level, it
is important to study this reduction at higher loop orders to confirm
its relation with integrability.

This question of higher loop orders was addressed recently in
\cite{Hoare:2019ark}, where we showed that, starting from 2 loops,
renormalizability requires a specific deformation of the classical target space
geometry, which may be interpreted as the result of adding
finite
local counterterms.

In \cite{Hoare:2019ark} we focused on the simplest examples of bosonic
integrable \sms{} with 2-dimensional target spaces. Here we shall consider
more general examples with higher-dimensional target spaces
and including $\BB$-field couplings.
We shall concentrate on a
particular class of integrable models: the $\l$-deformation based on a group
$G$ or a symmetric space $G/\HH$ (related to the coset \sm)
\cite{Sfetsos:2013wia,Hollowood:2014rla} with Lagrangian
\unskip\foot{\ Our notation and conventions are summarized in Appendix A.
In particular, we use hermitian generators $T^a$ of the Lie algebra so that if $g= e^v \in G$ then $v= i \, T_a v^a \in {\rm Lie}(G)$ is anti-hermitian.
The action is defined as $ {\cal S} = \tfrac{1}{4 \pi} \int d^2 \s \L$ so that $\L$
has extra factor of 2 compared to the ``conventional'' normalization.
}
\begin{align} \no
\L & = k \, \Big( {\PCM}(g) + {\WZ}(g) + \Tr\big[ J_+ A_- - A_+ K_- + g^{-1} A_+ g A_- - A_+ A_- - ( \lambda^{-1}-1) A_+ P A_-\big] \Big) , \\
& {\PCM} = - \ha \Tr[J_+ J_-] \ , \quad dB_{\rm WZ}=\tfrac{1}{6} \Tr [J\wedge J \wedge J] \ , \quad
J=g^{-1} dg \ , \quad K = dg \, g^{-1} \ , \la{1} \\
&\qquad P = \begin{cases}
1 \ , \quad \ \ \ \ \ \text{group space }G\\
P_{G/\HH} \ , \quad \text{symmetric space }G/\HH \ ,
\end{cases} \la{1111}
\end{align}
where $g \in G$, $A_\pm \in \Lie(G)$ and $P_{G/\HH}$ is the projector onto the orthogonal complement of $\Lie(\HH)$ in $\Lie(G)$.
Instead of $\l$ it is often convenient to use
the parameters $\gg$ or $\ka$ defined as
\begin{equation}\label{133}
\gg = \lambda^{-1}-1
\ , \qquad \qquad \ka = \frac{1-\lambda}{1+\lambda} \ .
\end{equation}
This ``\lam'' is special due to its close connection to the (gauged) WZW model
\begin{equation}\begin{split}\la{144}
\L_{G}(g) & = {\PCM}(g) + {\WZ}(g) \ ,
\\
\L_{G/H}(g,A) & =
{\PCM}(g) + {\WZ}(g) + \Tr\big[ J_+ A_- - A_+ K_- + g^{-1} A_+ g A_- - A_+ A_- \big] \ .
\end{split}\end{equation}
For example, the \lam{} for a group $G$ is a deformation of the $G/G$ gauged WZW model $ \L_{G/G} (g, A)$
by the term $ \gg A_+ A_-$.
This model is a particular $H=G$ case
of the one
considered in \cite{Sfetsos:1993bh,Tseytlin:1993hm}
\be\la{444}
\L = k \Big[ \L_{G/H}(g,A) - \gg \Tr (A_+ A_-) \Big]\ , \qquad \qquad g \in G \ , \qquad A_\pm \in \Lie(H) \ . \ee
This ``$\gg$-model'' \rf{444} interpolates
between two conformal theories: $G/H$ gauged WZW model (${\gg} =0$)
and $G/H$ chiral gauged WZW model (${\gg} =-1$) \cite{Chung:1992mj}.

Let us note also
that there is a $\mathbb{Z}_2$ transformation \cite{Itsios:2014lca,Hoare:2015gda} (see also \cite{Kutasov:1989aw,Tseytlin:1993hm})
\unskip\foot{\ Such a symmetry was discussed in a similar $\s$-model context in \cite{Tseytlin:1993hm} (see footnote 3 there).}
\begin{align}
& k \to - k \ , \qquad\qquad \l \to \l^{-1} \ , \qquad {\rm i.e. } \quad \ka \to -\ka \ , \label{z2sym}
\\ & g\to g^{-1} \ , \qquad A_+ \to A_+ - (1-\l^{-1}) P A_+ \ , \qquad A_- \to g A_- g^{-1} - K_- \ , \label{z2sym1}
\end{align}
that maps the Lagrangian \eqref{1} to itself.
The preservation of this symmetry at the quantum level may require a particular choice of
regularization scheme (see below).
Since the $\l \to 0$ (or $\gg \to \infty$) limit of the \lam{} yields a (gauged) WZW model,
we expect this to correspond to a fixed point of the RG flow.
The transformation \eqref{z2sym},\rf{z2sym1} then implies that the same should apply to the limit $\l \to\infty$ (or $\gg \to -1$).
Indeed, in the group space case the $\l \to \infty$ limit of \eqref{1} is conformal:
it is the $G/G$ chiral gauged WZW model, which, on
integrating out $A_\pm$, gives the $G$ WZW model at level $-k$.
Similarly, in the coset case we find in this limit the $G/H$ gauged WZW model at level $-k$.

\medskip

Integrating out the 2d gauge field $A_\pm$ in \rf{1}, i.e. reducing the model to the standard (or ``physical'') configuration space,
one finds a \sm{} with parameters $k$ and $\l$. The limit $\lambda \to 0$
yields the $G/\HH$ gauged WZW model (or $G$ WZW model in the group space case) with level $k$.
As in the examples in \cite{Hoare:2019ark}, we shall find that to preserve renormalizability of this model
at the 2-loop level with only the coupling $\l$ running, one must make a non-trivial modification of
the classical target space geometry.

At the same time, our central observation will be that, before integrating out $ A_\pm$,
the \lam{} is renormalizable without any deformation.
Changing the variables from $A_\pm$ to $h, \bar h \in G$ as
$A_+ = h \del_+ h^{-1}, \ A_- = \bar{h} \del_+ \bar h^{-1}$
gives a \sm{}
on the extended or ``tripled'' ($G \times G \times G$) configuration space $(g, h, \bh)$.
It may be interpreted as the sum of a decoupled $G$ WZW model and a deformation of the $G \times G$
WZW model by a particular left-right current interaction term.
In the group space case, the form of the resulting
action is then protected under the RG flow by the underlying chiral gauge symmetries
together with the global symmetries.

For the coset $G/\F$, the \lam{} is formally defined for any choice of $\F$
(with dependence on
the choice of $\F$ only through the projector $P$ in \rf{1}).
However, it is known to be integrable if $G/\F$ is a symmetric space \cite{Sfetsos:2013wia,Hollowood:2014rla}.
\unskip\foot{\ In this case it is also related to the
standard symmetric space \sm{} (which is both integrable and renormalizable) in the NAD limit \eqref{13}.}
We shall find evidence that the model is also renormalizable
if $G/\F$ is a symmetric space, which is a further indication of a connection between integrability and renormalizability.
\unskip\foot{\ For example, for more general cosets the
symmetries may not be sufficient to rule out other possible current-current counterterms not present in the classical action.}

The \lam{} \eqref{1} admits a special limit
$\l\to 1$, $k \to \infty$ with $\hh\equiv 2k (1-\l)$ fixed
\begin{align}\label{13}
&\lambda = 1-\ha \hh k^{-1} + \O(k^{-2}) \ ,
\qquad \ \qquad k\to \infty \ , \\
& g = \exp(- \ha \hh k^{-1} v)= 1 - \ha \hh k^{-1} v + \O(k^{-2}) \ , \qquad\qquad \ v \in \Lie(G) \ ,\no
\end{align}
resulting in the following first-order Lagrangian \cite{Sfetsos:2013wia}
\begin{equation}\label{14}
\L = \tfrac{1}{2} \hh\,\Tr\Big[v \big(\partial_+ A_- - \partial_- A_+ + [A_+,A_-]\big) - A_+ P A_-\Big] \ .
\end{equation}
This is an interpolating model for non-abelian duality:
integrating out $v$ in \eqref{14} gives the
principal chiral model (PCM) on group $G$, or the $G/\F$ symmetric space
$\sigma$-model,
with coupling $\sim \hh^{-1}$, while integrating out $A_\pm$ gives the corresponding non-abelian dual (NAD) model.
The renormalizability of the \lam{} in the extended configuration space also applies in this
limit: although the NAD of a group or symmetric space \sm{}
requires a non-trivial deformation at the 2-loop level \cite{Hoare:2019ark},
the interpolating model remains renormalizable without deformation.
We conclude that, staying at the level of the interpolating model,
non-abelian duality commutes
with the RG flow beyond the 1-loop level
(thus resolving problems discussed in \cite{n1,n2}).

\medskip

To study the 2-loop renormalizability of the above models we will
be using the explicit expression for the $\b$-function of the general bosonic \sm
\begin{equation}\begin{split}
\mathcal{S} \equiv \frac{1}{4\pi}\int d^2 \sigma \, \mathcal{L} &= - \frac{1}{4\pi}\int d^2 \sigma \, \big(\eta^{rs} \GG_{\mu\nu}(x) + \epsilon^{rs} \BB_{\mu\nu}(x)\big)\partial_r x^\mu \partial_s x^\nu \la{15}
\\
&= \frac{1}{4\pi} \int d^2\sigma \, \big(\GG_{\mu\nu}(x) + \BB_{\mu\nu}(x)\big)\partial_+ x^\mu \partial_- x^\nu \ .
\end{split}\end{equation}
In terms of the curvature ${\hat{R}^\m}{}_{\n\r\s}$ of the generalized connection $\hat\Gamma^\mu{}_{\nu\rho} = \Gamma^\mu{}_{\nu\rho}(\GG) - \tfrac12 \H^\mu{}_{\nu\rho}$
the 2-loop RG equation can be written as
\cite{Friedan:1980jm,Braaten:1985is,mt,Hull:1987pc}
\begin{equation}\begin{split}\la{rg}
\frac{d(\GG_{\mu\nu}+\BB_{\mu\nu})}{dt} & + L_\XX (\GG+\BB)_{\mu\nu} + (dY)_{\mu\nu} = \beta_{\mu\nu}^{(1)} + \beta_{\mu\nu}^{(2)} + \ldots
\\ & = \hat R_{\mu\nu} + \tfrac12 \Big[\hat R^{\rho\sigma\tau}{}_\nu \hat R_{\mu\rho\sigma\tau} - \tfrac12 \hat R^{\sigma\tau\rho}{}_\nu \hat R_{\mu\rho\sigma\tau}
+\tfrac12 \hat R_{\rho\mu\nu\sigma} (\H^2)^{\rho\sigma} \Big] + \ldots \ .
\end{split}\end{equation}
Here
$t$ is log of the RG mass scale,
$L_X$ is the Lie derivative with respect to the vector $X$, corresponding to RG-dependent diffeomorphisms, and $dY$ is an exact two-form, which is a total derivative when pulled back to the worldsheet.
This 2-loop $\b$-function is given in a particular ``minimal'' subtraction scheme \cite{mt}.
\unskip\foot{\ Alternative ``minimal'' schemes are related to this one by $(\GG+\BB)_{\mu\nu} \to (\GG+\BB)_{\mu\nu} + a_1 R_{\mu\nu} + a_2 (\H^2)_{\mu\nu} + a_3 \hat R_{\mu\nu}$. Since $\beta_{\mu\nu}^{(1)} = \hat R_{\mu\nu}$ it follows that shifts by $\hat R_{\mu\nu}$ will leave $\beta_{\mu\nu}^{(2)}$ invariant. On the other hand, shifts by $ R_{\mu\nu}$ and $(\H^2)_{\mu\nu}$ do modify $\beta_{\mu\nu}^{(2)}$, and hence, in the case of non-trivial $\BB$-field, the 2-loop RG equation is no longer scheme-independent \cite{mt}. One other scheme that will be useful in our discussion of T-duality in section \ref{standard} is related to
the minimal one in \rf{rg} by $(\GG+\BB)_{\mu\nu} \to (\GG+\BB)_{\mu\nu} + \tfrac14 (\H^2)_{\mu\nu}$.}

\bigskip

Let us now comment on the motivation behind the present work. In
addition to understanding non-abelian duality beyond the 1-loop level,
investigating the $\lambda$-model and its quantum corrections is of more
general interest in the context of integrable deformations of superstring
actions in special $AdS$-type backgrounds. Integrability has been a powerful
tool in the proposed solution of the spectral problem for string theory in on
$AdS_5 \times S^5$ dual to the large-$N$ maximally supersymmetric YM
theory \cite{adscftsol}. This motivates the study of further similar models,
potentially leading to new exact solutions of strings in curved spaces and
dual gauge theories. By now there are many examples, including those based on
lower-dimensional $AdS$ spaces \cite{Zarembo:2010sg,Wulff:2014kja}, as well as
deformed backgrounds, such as the well-studied $\beta$-deformation
\cite{betadef}.

The $\lambda$-deformation of the $AdS_5 \times S^5$ superstring
\cite{Hollowood:2014qma} belongs to a more general class of integrable
deformations not obtained by T-duality. It is a deformation of the non-abelian
dual model of the undeformed superstring model and is closely related to the
$\eta$-deformation \cite{Delduc:2013qra}, which is a deformation of the
superstring action itself. The latter generalises the bosonic $\eta$-model of
\cite{klimcik,dmv}. More precisely, the $\lambda$-model and $\eta$-model are
related by the Poisson-Lie duality \cite{poissonlie} (which is a generalisation
of non-abelian duality) and a particular analytic continuation
\cite{Hoare:2015gda,pld,hs}. While both models describe a string propagating
in a type II supergravity background \cite{sugra}, much remains to be
understood about their structure. Probing the quantum properties of the
bosonic $\eta$-model and $\lambda$-model (even though they are not themselves
scale-invariant theories suitable for defining string models) can provide
valuable insights into their superstring counterparts. For example, the
relation via the Poisson-Lie duality was first understood in the bosonic case.
It is thus natural to first explore the question of quantum corrections by
studying the bosonic models.

It is also worth emphasizing that the bosonic models are of interest in their
own right in the context of investigation of general integrable 2d theories.
The $\eta$-model has played an important role in generalizing the duality
between the deformed $O(3)$ and $O(4)$ sigma models and massive integrable QFTs
\cite{Fateev:1992tk} to higher-rank groups \cite{Fateev:2019xuq}. While the
dual theories are quantum-exact, the \sm{} side of the duality is only
understood so far to leading order in the loop expansion. Therefore, after
finding quantum corrections to integrable \sms{} consistent with
renormalizability, studying their compatibility with this duality may be a
useful way to further explore the conjectured relationship between
integrability and renormalizability.

\bigskip

The rest of the paper is organized as follows.
In section \ref{extended} we consider the $\lambda$-model on the extended configuration space $(g,A_+,A_-)$.
We argue that it should be renormalizable with only the parameter $\l$ running and
compute
the 2-loop $\b$-function of
the $\lambda$-model based on general groups and symmetric spaces.
As a consequence, the same renormalizability conclusion holds
for the model \rf{14} interpolating between the PCM
and its NAD, with the same 2-loop $\beta$-function for $\hh$ as in the PCM
(and the same in the symmetric space case).

In section \ref{standard} we study the renormalization of the $\l$-model
defined by the standard \sm{} action, after integrating out $A_\pm$.
In this case its invariance under the 2-loop RG flow
requires a specific deformation of the classical geometry. While in the $SU(2)/U(1)$ case
the required counterterm is simple \cite{Hoare:2019ark}, in the $SU(2)$ case,
the corresponding quantum-corrected \sm{} action has a rather intricate structure. We also consider a particular limit of the $SU(2)$ \lam{}
where
it becomes T-dual to a \sm{} for a squashed 3-sphere,
explaining the consistency of the quantum deformation of the original \lam{} with the known quantum correction
to the T-duality transformation rule. We also discuss the 2-loop $\beta$-function for the NAD of the $SU(2)$ PCM.

Some concluding remarks are made in section \ref{conclusions}.
In Appendix \ref{A} we summarize our notation and group-theory conventions.
In Appendix \ref{B} we compute the 2-loop $\b$-function for a two-coupling
``squashed'' principal chiral model that interpolates between the $G$ group space PCM
and the $G/\F$ coset \sm, determining also the 2-loop
$\b$-function
for the latter.

\section{Renormalizability of \texorpdfstring{$\l$}{lambda}-model: extended configuration space}\label{extended}

In this section we shall study renormalization of the $\lambda$-model on the extended configuration space. It is first useful to draw an analogy with
the $\gg$-model \cite{Sfetsos:1993bh} defined in \rf{444} (where $G$ and $H\subset G$ are simple Lie groups) that
interpolates between the gauged WZW (gWZW) and chiral gauged WZW (cWZW) theories.
Changing variables $(A_+, A_-) \to (h, \bar{h})$ as
\be\la{21}
A_+ =h\, \del_+ h^{-1} \ , \qquad A_- = \bar{h}\, \del_- \bar{h}^{-1} \ , \qquad \qquad h,\ \bar{h} \in H \ ,
\ee
and using the Polyakov-Wiegmann identity \cite{Polyakov:1983tt},
we can rewrite the Lagrangian in \eqref{444} as a combination of
WZW models (cf. \rf{144})
\begin{equation}\begin{split}\la{22}
\L &= k \Big( \L_{G/H}(g,A) - {\gg} \Tr [A_+ A_-] \Big) \\
&= k \Big( \L_G (h^{-1} g \bar{h}) - \L_H ( h^{-1} \bar{h} ) - {\gg} \big[ \L_H (h^{-1} \bar{h}) - \L_H ( h^{-1} )- \L_H (\bar{h} )\big] \Big) \ .
\end{split}\end{equation}
The change of variables
in \rf{21} results in a Jacobian contributing to the action as
\cite{Polyakov:1983tt,Polyakov:1988qz}
\begin{align}
\Delta \L & = - 2 c_{_H} \Big( \L_H(h^{-1} \bar{h} ) + \q \Tr [A_+ A_-]\Big)\no \\
&= - 2 c_{_H} \Big( \L_H(h^{-1}) + \L_H ( \bar{h} ) + ( \q-1) \Tr [ \del_+ h \, h^{-1} \, \del_- \bar{h} \, \bar{h}^{-1} ]\Big) \ , \la{jac}
\end{align}
where $c_{_H}=c_2(H)$ is the dual Coxeter number of $H$ and
an arbitrary coefficient $\q$ parametrizes the ambiguity of adding a local counterterm $\Tr [A_+ A_-]$.
Combining \eqref{jac} and the classical Lagrangian \eqref{22} gives
\be
\L = k \L_G (h^{-1} g \bar{h}) - (k+2 c_{_H}) \L_H ( h^{-1} \bar{h} ) - (k {\gg} + 2 \q c_{_H}) \big[ \L_H (h^{-1} \bar{h}) - \L_H ( h^{-1} )- \L_H (\bar{h} ) \big] \ . \ \la{5term}
\ee
In the special cases ${\gg} =0$ (gWZW model) and ${\gg} = -1$ (cWZW model), we can choose q such that \eqref{5term} is a sum of WZW models \cite{Sfetsos:1993bh}
\begin{align}
&{\rm gWZW}: \ \ && {\gg} =0, && \q=0, && \L = k \L_G (h^{-1} g \bar{h}) - (k+2 c_{_H}) \L_H ( h^{-1} \bar{h} ) \ , \la{gWZW} \\
&{\rm cWZW}: \ \ && {\gg}=-1, \ \ && \q=-1, \ \ && \L = k \L_G (h^{-1} g \bar{h}) - (k+2 c_{_H}) \big[ \L_H ( h^{-1} ) + \L_H (\bar{h} ) \big] \ . \la{cWZW}
\end{align}
Since the arguments $\{\td g = h^{-1} g \bar{h},\ \td h = h^{-1} \bar{h}\}$ in \eqref{gWZW}
or $\{\td g = h^{-1} g \bar{h}, \ h^{-1},\ \bar{h}\}$ in \eqref{cWZW}
may be treated as independent fields,
it follows that these two models are conformally invariant.

For general values of $\g$, choosing $\q=-1$ 
as in the cWZW case \rf{cWZW},
we may rewrite \rf{5term} as
\begin{align}
&\L= k \L_G (\tilde{g}) + \L'(\tih, \bh) \ ,\qquad \qquad \qquad \tilde{g} \equiv h^{-1} g \bar{h}\in G \ , \qquad \qquad \tih\equiv h^{-1} \in \F \ , \la{277} \\
&\L'= - (k+2 \cf) \big[ \L_\F ( \tih) + \L_\F (\bar{h} ) \big] + k(1+{\gg}) \Tr [ \tih^{-1} \del_+ \tih \, \del_- \bh \bh^{-1}] \ .
\la{27}
\end{align}
This model is defined on the extended configuration space $(\tilde{g},\tih,\bar{h}) \in G \times \F \times \F $. The
first $G$ WZW term (which is conformal on its own)
decouples and then we are left with the ``truncated'' model $\L'$ on $\F \times \F $
which is simply a sum of two group $\F$ WZW models perturbed by the product of the left and right currents.
Like the chiral gauged WZW model \rf{cWZW} the Lagrangian $\L'$ in \rf{27} is invariant
under the chiral gauge symmetry
$\tih \to u(\s^-)\, \tih , \ \bh \to \bh\, w(\s^+) , \ u, w \in \F$ as well as the global $\F$ symmetry
$\tih \to \tih v_0,\ \bh \to v_0^{-1} \bh, \ v_0 \in \F$.
As we shall argue for the $\l$-model,
these two symmetries imply that the $\g$-model is also renormalizable with only one coupling $\gg$ running.

Let us note that the $\gg$-model \rf{444},\rf{22} admits also a generalization similar to the coset case of the \lam{} in \rf{1}:
with $ {\gg} A_+ A_- $ term replaced by $ {\gg} A_+ P A_- $, where $P$ is the projector onto the $\HH/F$ coset part of the algebra of $\HH$
(with $F \subset \F \subset G$). When $H/F$ is a symmetric space this model should again be renormalizable
on the extended configuration space $(\tilde{g}, \tih,\bh) \in G\times \F\times \F$
due to chiral gauge symmetry.

\subsection{Group space}

Let us now apply similar arguments to the $\l$-model for the group $G$, which
is given by \rf{1} with $P=1$. Taking $H = G$ in \eqref{21}, so that now
$h,\bar h \in G$, we obtain \eqref{5term} with $H=G$ and $c_{_H} \to \cg = c_2(G)$.
It will represent the $\l$-model
as a \sm{} on a ``tripled'' configuration space $(\td g = h^{-1} g \bar{h}, \, h^{-1},\, \bar{h}) \in G\times G\times G$.
Since the $\q$-dependent term in \eqref{jac} is simply equivalent to a finite quantum (order $1/k $) redefinition
${\gg} \to {\gg} + \tfrac{ 2 \cg}{ k}\q $ of the parameter $\g$,
we are free to fix $\q$ to a specific value, $\q=-1$,
as in the cWZW case \eqref{cWZW} and in \rf{277},\rf{27}.
This gives the following analog of \rf{277}
\begin{align}\no
\L & = k \L_G (\tilde{g}) - k(1+{\gg}) \L_G ( h^{-1} \bar{h} ) + (k {\gg} - 2 \cg) \big[ \L_G ( h^{-1} ) + \L_G (\bar{h} )\big] \\
\la{effective}
& = k \L_G (\tilde{g}) - (k+2 \cg) \big[ \L_G ( h^{-1} ) + \L_G (\bar{h} ) \big] - k(1+{\gg}) \Tr [ \del_+ h h^{-1} \, \del_- \bh \bh^{-1}] \ .
\end{align}
We thus obtain the same tripled theory as \rf{277},\rf{27}, now with $\F\to G$: the first term is the $G$ WZW model for $\tilde{g}$, which decouples from the
$(\hat h, \bar h)$ theory described by the ``truncated'' Lagrangian
\begin{align} \la{6dmodel}
& \L'(\tih, \bar h)
= - \tk \Big( \L_G ( \tih ) + \L_G (\bar{h} ) - \tl \Tr [ \tih^{-1}\del_+ \tih \, \del_- \bh \bh^{-1} ] \Big) \ , \qquad \qquad \tih \equiv h^{-1}, \ \bh \in G\ , \\
& \ \ \ \label{redef}
\tk \equiv k + 2 \cg \ ,\qquad \qquad \tl \equiv \frac{k}{k+2\cg}(1+{\gg})
= \frac{k}{\tk}\l^{-1} = \l^{-1} + \O (k^{-1}) \ .
\end{align}
$\L'$ may be interpreted as the Lagrangian for
the two WZW models for the two groups $G $
with the same level $-(k+ 2 \cg)$
perturbed by a product of the left current of one group and the right current of the other.
Classical integrability of the model \rf{6dmodel} (implying integrability of the group space $\l$-model) 
was first shown in \cite{Bardakci:1996gs}.
\unskip\foot{\ Similar models were discussed in the past, e.g., in \cite{Guadagnini:1987ty,Hull:1995gj}.
A generalization of this model was also considered in
\cite{Georgiou:2016urf,Georgiou:2017aei,Georgiou:2017oly,Georgiou:2017jfi},
where it was interpreted as a special case of a ``doubly $\l$-deformed'' $\s$-model.
Our path-integral relation
between the $\l$-model \rf{1} and the truncated model \rf{6dmodel}
should be equivalent (at least at the classical and 1-loop level)
to the canonical equivalence between the doubly $\l$-deformed model
and two copies of the $\l$-model found in \cite{Georgiou:2017oly}
(upon setting one of the two $\l$-parameters to zero).
The leading order in $1/k$ (1-loop) renormalization of similar models
was studied earlier in \cite{Gerganov:2000mt,LeClair:2001yp}.}

Our central observation is that the model \eqref{6dmodel}
is renormalizable with only
the coupling $\l$ (or $\tl$) running with the RG scale ($k$ should not run as it appears as the coefficient of
the WZ term).
Indeed, the structure of
\eqref{6dmodel} is protected by the same chiral gauge symmetry
present in the cWZW model \eqref{cWZW} and in the $\gg$-model \rf{27},
\be \la{210}
\tih \to u(\s^-)\, \tih \ , \qquad \bh \to \bh\, w(\s^+) \ , \qquad \qquad u, w \in G\ . \ee
This symmetry, together with the global $G$ symmetry
\be \tih \to \tih g_0 \ , \qquad \bh \to g_0^{-1} \bh \ , \qquad\qquad g_0 \in G \ , \ee
prohibits the appearance of other current-current interaction terms under the RG flow.
\unskip\foot{\ The presence of this symmetry is also a manifestation of the integrability of the original $\l$-model.}

It is then straightforward to compute the $\beta$-function for $\l$
in the large $k$ perturbation theory, which we will do in section \ref{groupbeta}.
The two fixed points of the RG flow for \eqref{6dmodel} will be $\tl = \infty,0$, corresponding to $\l=0,\infty$ respectively.
\unskip\foot{\ \label{f1} Note that for general values of $\q$ in \eqref{jac},\eqref{5term} one gets
$\tk \tl = k \l^{-1} + 2 (1+ \q) \cg$
in \rf{redef}.
Finite redefinitions of parameters like $k \to \tk$ and $\l\to \tl$ in \eqref{redef}
are not important for the discussion of renormalization
in $1/k$ perturbation theory,
simply
reflecting the freedom of scheme choice.
They may, however, correct the 1-loop fixed points $\l=0,\infty$ of the RG flow.}

\subsection{Coset space}

Next, let us consider the $\lambda$-model for the
coset $G/\F$, i.e. setting $P=P_{G/\F}$ in \eqref{1}.
Repeating the same steps, i.e.
using \eqref{21} to introduce $h,\bh\in G$,
including the contribution from the Jacobian \eqref{jac},
setting $\tilde{g} = h^{-1} g \bar h$ and $ \tih=h^{-1}$, we get
\be\label{effectivecoset1}
\L = k \L_G (\tilde{g}) - (k+2\cg) \L_G ( \tih\bar{h} ) + k {\gg} \Tr [ \tih^{-1}\del_+ \tih P\del_- \bh \bh^{-1} ]
+ 2 \q \cg \Tr [ \tih^{-1}\del_+ \tih \, \del_- \bh \bh^{-1} ]\ . \ \ \
\ee
Classically (i.e. for large $k$)
this model has the
expected $\F$ gauge symmetry
\begin{equation}\label{gsym}
\tih \to \tih f \ , \qquad \bar h \to f^{-1} \bar h \ , \qquad\qquad f=f(\s^+,\s^-) \in \F \ .
\end{equation}
To preserve this gauge symmetry in \eqref{effectivecoset1}
let us choose (as in the gWZW case \eqref{gWZW}) $\q=0$, thus obtaining
\begin{equation}\begin{split}\label{effectivecoset2}
\L & = k \L_G (\tilde{g}) - (k+2\cg) \L_G ( \tih\bar{h} ) + k {\gg} \Tr [ \tih^{-1}\del_+ \tih \, P\, \del_- \bh \bh^{-1} ] \ ,
\\ &= k \L_G (\tilde{g}) - (k+2\cg) \big[\L_G ( \tih)
+ \L_G(\bar{h} )\big] + \Tr [ \tih^{-1}\del_+ \tih\, (k+2\cg + k {\gg} P)\, \del_- \bh \bh^{-1}] \ .
\end{split}\end{equation}
As in
\eqref{effective}
the
WZW term for the field $\tilde{g}$ decouples,
leaving us with
\begin{align}
&\L'(\tih, \bh)= -\tk \Big( \L_G ( \tih\bar{h} ) - (\tl-1) \Tr [ \tih^{-1}\del_+ \tih\, P\, \del_- \bh \bh^{-1} ] \Big)\no \\
& =- \tk \Big(\L_G ( \tih) + \L_G(\bar{h} ) - \Tr[\tih^{-1}\del_+\tih\, (1-P)\, \del_-\bh\bh^{-1} ] -
\tl \Tr [ \tih^{-1}\del_+ \tih P\del_- \bh \bh^{-1}] \Big), \label{214}
\\
& \label{215}
\qquad \tk \equiv k + 2 \cg \ , \qquad \tl \equiv 1+ \frac{k {\gg}}{k+2\cg}
=\frac{k}{\tk}\l^{-1} +\frac{2 \cg}{\tk}
= \l^{-1} + \O (k^{-1} ) \ .
\end{align}
$\L'$ represents the $G \times G$ WZW model deformed by the product of the left and
right currents projected to the subgroup $P_\F= 1 -P$ and the coset $P_{G/\F}=P$.
The classical integrability of the model \rf{214} (implying also the integrability of the $G/H$ $\l$-model)
was argued in \cite{Bardakci:1996gs} (example 4 there).

While it is again invariant under the chiral transformations \eqref{210},
here it is not immediately clear that the theory \rf{214} is renormalizable with only one coupling running:
in principle, different gauge-invariant projections of the product of currents may appear as independent counterterms.
When $G/\F$ is an irreducible symmetric space (which also implies the
integrability of the \lam),
the coset part of the algebra of $G$ transforms in an irreducible representation of $\Lie(H)$.
Thus
the model is renormalizable with only $\l$ running since new current-current interaction terms are prohibited by symmetries.
\unskip\foot{\ If the coset directions $G/\F$ transform in an irreducible representation of $\Lie(\F)$ that is reducible over $\mathbb{C}$,
i.e. decomposes into two complex conjugate representations,
then there can be additional real gauge-invariant terms that respect the chiral symmetry.
An example would be $i\Tr [ \tih^{-1}\del_+ \tih (P_+ - P_-)\del_- \bh \bh^{-1}]$
where $P = P_+ + P_-$ and $P_\pm$ are projectors onto the conjugate representations.
Such a term should not be generated under the RG flow as it is not invariant under $\tih \leftrightarrow \bh^{-1}$ plus parity, which is a symmetry of \eqref{214}.
More generally, we expect any new terms to be excluded by symmetries.\label{caveat}}
We shall see this explicitly at the 2-loop level in section \ref{cosetbeta} below.

In this case the two expected fixed points of the RG flow for \eqref{214}, $\tl^{-1} = 0$ and $\tl^{-1} = \infty$, now correspond to $\l = 0$ and $\l = - \frac{k}{2c_{_G}}$ due to the shift in \eqref{215}.
That is, one of the 1-loop fixed points, $\l = \infty$, is corrected.

\subsection{2-loop \texorpdfstring{$\b$}{beta}-function of \texorpdfstring{$\lambda$}{lambda}-model for
group \texorpdfstring{$G$}{G}} \la{groupbeta}

Let us now compute the 2-loop $\b$-function for the model \eqref{6dmodel}
in the case that $G$ is a compact simple group, explicitly
demonstrating its renormalizability with only one parameter $\l$ running
(the 1-loop $\b$-function for this
$G\times G$
model was computed in
\cite{Georgiou:2017aei,Georgiou:2017jfi}).
\unskip\foot{\ On the standard configuration space the
1-loop $\b$-function of the \lam{} for group space $G$ or symmetric space $G/\F$
can be extracted from \cite{Tseytlin:1993hm} and was also explicitly computed
in \cite{Sfetsos:2014jfa,Appadu:2015nfa}.}
We shall use large $k$ perturbation theory with $\l$ arbitrary.

Let us introduce the basis $\{T^a\}$ for $\Lie(G)$ (see Appendix A for conventions).
In a slight abuse of notation, we shall use the indices $a$ and $\bar{a}$ for the two
$G$-valued fields $\tih$ and $\bar{h}$ in \eqref{6dmodel} respectively (with the tangent space index for $G \times G$ denoted as
$A = \{a,\bar a\}$). As in
\cite{Georgiou:2017jfi} we introduce the vielbein
\begin{align}
&E^A = (e^a, e^{\bar{a}}) = (\sqrt{1-\tl^2}\, \RR^a,\, \LL^{\bar{a}} + \tl \RR^a) \ , \la{vielbein}\\
&\RR^a \equiv \Tr [ T^a \tih^{-1} d \tih] \ , \qquad \qquad \LL^{\bar{a}} \equiv \Tr [ T^a d \bar{h} \bar{h}^{-1}]\ ,
\end{align}
where $\RR^a$ and $\LL^a $ are the
currents that appear in the deformation term in \eqref{6dmodel}.
Up to permutations, the non-zero components of the metric and $\H$-tensor of the $G\times G$ model \eqref{6dmodel} are (cf. \rf{15})
\unskip\foot{\ Note that in our conventions (with hermitian generators $T^a$, see Appendix A)
the vielbein defined in \rf{vielbein} and the components $\H_{ABC}\sim f_{ABC}$ in \rf{Hflux} are imaginary but
the 3-form $\H = \tfrac{1}{6} \H_{ABC} E^A \wedge E^B \wedge E^C$ is real.}
\begin{gather} \GG_{ab} = \GG_{\bar{a}\bar{b}} = \tfrac{\tk}{2} \d_{ab} \ , \la{vielmet}
\\
\H_{abc} = -\tfrac{\tk}{2} \tfrac{\sqrt{1-\tl^2}(1+2\tl)}{(1+\tl)^2} f_{abc} \ , \qquad
\H_{\bar{a}\bar{b}\bar{c}} = -\tfrac{\tk}{2} f_{abc} \ , \qquad
\H_{\bar{a}bc} = -\tfrac{\tk}{2} \tfrac{\tl}{1+\tl} f_{abc} \ . \la{Hflux}
\end{gather}
Our aim is to compute the corresponding $\b$-function in \rf{rg}.
Let us formally define the torsion as
$T^A = \ha {\H^A}_{BC} E^B \wedge E^C $ where the tangent space index is raised with the inverse of the metric \eqref{vielmet}.
Then from the Cartan structure equation
$d E^A + {\hat{\w}^A}{}_{B} \wedge E^B = T^A$, we obtain the torsionful spin connection
${{\hat \w}^A}{}_{B} = {\hat{\w}^A}{}_{BC} E^C$ with non-zero components
\be
(\tl^{-1}+1){{\hat \w}^a}_{ \ b} =
\hat \w^{\bar{a}}{}_{\bar{b}} = f^{a}{}_{bc} \big(- \tfrac{\tl}{\sqrt{1-\tl^2}} e^c + e^{\bar{c}} \big) \ .
\ee
The non-zero components of the curvature,
${\hat{R}^A}{}_{B} = d{\hat{\w}^A}{}_{B} + {\hat{\w}^A}{}_{C} \wedge {\hat{\w}^C}{}_{B}=\ha {\hat{R}^A}{}_{BCD} E^C \wedge E^D$, are then found to be
\begin{equation}\begin{split}\la{torsionriem}
(\tl^{-2}-1){\hat{R}^a}{}_{bde} =
-\sqrt{\tl^{-2}-1}{\hat{R}^a}{}_{b d \bar{e} }
= {\hat{R}^a}{}_{b \bar{d} \bar{e} } = \frac{\tl}{(1+\tl)^2} f^a{}_{bc} f^c{}_{de} \ .
\end{split}\end{equation}
Plugging \rf{torsionriem} into \eqref{rg} one obtains
\begin{equation}\la{220}
(\beta^{(1)}_{\m\n} + \beta^{(2)}_{\m\n} )\, dx^\m \otimes
dx^\n = \sum^{{\rm dim}\, G}_{a=\bar a=1}\Big[ \frac{2 \cg \tl^2}{(1+\tl)^2}
+ \frac{4 \cg^2}{\tk} \frac{\tl^4(1-2\tl)}{(1+\tl)^5(1-\tl)} \Big] \RR^a \otimes \LL^{\bar{a}} \ .
\end{equation}
Here $\otimes$ indicates that the product is not symmetrized.
We conclude that only the $\tl$-dependent term in \eqref{6dmodel} gets renormalized, i.e.
the 2-loop RG equation \eqref{rg} is solved with $X^\mu=Y_\mu=0$ and
\be\label{01}
\ddt \tk =0 \ ,\qquad \qquad \ddt \tl = \frac{2 \cg}{\tk} \Big[ \frac{\tl^2}{(1+\tl)^2} + \frac{2 \cg}{\tk} \frac{\tl^4(1-2\tl)}{(1+\tl)^5(1-\tl)} \Big] \ .
\ee
Here the 1-loop term agrees with \cite{Sfetsos:2014jfa,Appadu:2015nfa} (recall that $\tilde{\l} = \l^{-1} + \ldots$ and $\tilde{k} = k + \ldots$, cf. \rf{215}).
Note also that, while the Lagrangian \rf{6dmodel} is linear in $\tilde \l$,
the non-polynomiality of \rf{01} in $ \tilde \l$ is a direct consequence of the exactness of the \sm{} $\b$-function \rf{rg} in the metric $\rm G$.

The level $k$ is thus RG-invariant as it should be and, as expected, $\tl = \infty,0$ are fixed points of the RG flow.
Expressing $\tk$ and $\tl$ in \eqref{redef} in terms of $k$ and the coupling
$\ka$ using \eqref{133}, we find
\be\ddt k =0 \ ,\qquad \qquad
\ddt \ka = \frac{\cg}{4k}(1-\ka^2)^2 \Big[ 1 + \frac{\cg(1-\ka)^2 (1 -10\ka - 3 \ka^2)}{8k \ka} \Big] \ . \la{24}
\ee
At the fixed point $\ka=1$
(equivalent to $\l=0$ or $\tl = \infty$), the \lam{} reduces to the $G$ WZW model with level $k$.
The other fixed point $\ka=-1$ (equivalent to $\l=\infty$ or $\tl=0$)
is the $G/G$ cWZW model, which reduces to the $G$ WZW model with level $-k$ after integrating out the gauge field.
\unskip\foot{\ \la{ftpcm} To compare,
for the PCM model plus WZ term $\L= \hh \PCM + k \WZ$
(which becomes the WZW model for $\k\equiv \frac{\hh}{k} =1$) the 2-loop RG equation
for $\k$
\cite{Witten:1983ar,mt,Bos:1987mw} is
$
\ddt \k = \frac{\cg}{k}(1-\k^{-2}) \big[ 1 + \frac{\cg}{2k\, \k}(1-3 \k^{-2} ) \big].
$} \unskip${}^{,}$
\unskip\foot{\ Taking
$\ka=0$ or $\l=1$ with finite $k$ (in contrast to the NAD limit
\eqref{13}) gives the $G/G$ gWZW model
in which one can fix a gauge so that the remaining degrees of
freedom correspond to the Cartan torus.}

Note that the 2-loop term in \eqref{24} is scheme-dependent: it can be changed by
redefining $\ka$ by a $1/k$ term (or shifting $k$ by a finite term).
\unskip\foot{\ For example, the
2-loop term in \eqref{01} will depend on the parameter $\q$ in
\eqref{jac},\eqref{5term}, cf. footnote \ref{f1}.}
Even though $k$ is not running, we effectively have a 2-coupling theory
(with $1/k$ playing the role of a loop-counting parameter)
so only the 1-loop term in the $\b$-function \rf{24} is scheme-independent.

In a general scheme, the symmetry $k \to - k, \ \l \to \l^{-1}$ \rf{z2sym}
of the 1-loop RG equation in \rf{01},\rf{24} is not manifest 
at the 2-loop level.
To preserve it requires a particular formulation of the quantum theory, i.e.
a specific definition of the couplings, or choice of scheme.
For example, if we redefine the parameters as
\begin{align} \la{2777}
(\tk, \tl) \to (\kk, \bl) \ , \qquad \qquad \tk =\kk + \cg\ , \qquad \quad \tl=\bl\Big[ 1 + \cg \kk^{-1} \frac{1-\bl}{1 +\bl} \Big]
\ , \end{align}
then the RG equation for $\bl$ resulting from \rf{01} is
\be \la{2789}
\frac{d}{dt}\bl = \frac{2 \cg}{\kk}\frac{\bl^2}{(1+\bl)^2}\Big[1 - \frac{2\cg}{\kk} \frac{\bl^2}{(1+\bl)^2} \frac{\bl+\bl^{-1}-1 }{1-\bl^2}
\Big] \ ,
\ee
which is invariant under
the following quantum version of \rf{z2sym}
\be\la{227}
\kk \to -\kk \ , \qquad \qquad \bl \to \bl^{-1} \ .\ee
Note that here $\kk=k+ \cg$ is the same as the shifted level of the WZW theory.
It is also worth observing that \rf{2777} is not the only redefinition that
restores the symmetry \eqref{z2sym}. 
Indeed, even restricting to those that preserve the
existence of the fixed points and the NAD limit discussed below, there are many.

In the NAD limit \eqref{13} (i.e. $\ka= \tfrac14 \hh k^{-1}, \ k\to \infty$)
the 2-loop RG equation \eqref{24} becomes
\be
\ddt \hh = {\cg} + \ha {\cg^2}\hh^{-1} \ . \la{25}
\ee
Here the 2-loop term is scheme-independent since this is now a 1-coupling theory.
As verified in Appendix \ref{B}, the $\beta$-function \eqref{25} matches the standard expression
(see, e.g., \cite{McKane:1979cm,Hikami:1980hi,Friedan:1980jm} and footnote \ref{ftpcm})
for the 2-loop $\b$-function of the PCM on a compact simple group $G$ (with the coupling $\hh= 2 {\rm g}^{-2}$).
As in the $SU(2)/U(1)$ example discussed in \cite{Hoare:2019ark}, this demonstrates that the NAD of the PCM
has the same 2-loop $\b$-function as the PCM itself, extending the previous conclusions \cite{Fridling:1983ha,Fradkin:1984ai}
on the 1-loop quantum
equivalence of models related by the non-abelian duality to the 2-loop level.

\subsection{2-loop \texorpdfstring{$\b$}{beta}-function of \texorpdfstring{$\lambda$}{lambda}-model for
symmetric space \texorpdfstring{$G/\F$}{G/\F}}\la{cosetbeta}

We now turn to the case of the \lam{} based on the compact irreducible symmetric space $G/\F$, explicitly demonstrating its renormalizability by computing the 2-loop $\beta$-function for the model \eqref{214}.
The computation runs analogously to the group space case discussed in section \ref{groupbeta}.
We decompose the basis $\{T^a\}$ of $\Lie(G)$ according to the $\mathbb{Z}_2$ grading of the Lie algebra with $\{T^\a\}$ spanning $\Lie(\F)$ and $\{T^i\}$ its complement (cf. Appendix \ref{A}).
\unskip\foot{\ In addition to the identities in Appendix A
here we have $f_{\a \b i} = f_{ijk} =0$ from the $\mathbb{Z}_2$ grading of $\Lie(G)$ (see \rf{a3}).}
The tangent space index for $G \times G$ now splits as $A = \{ a; \bar{a} \} =\{\a, i; \bar{\a}, \bar{\imath} \}$, with the unbarred and barred indices corresponding to the two $G$-valued fields $\hat h$ and $\bar{h}$.

Here we use the vielbein
\be
E^A = (e^i, e^{\bar{\imath}}, e^\a, e^{\bar\a}) = (\sqrt{1-\tl^2} \RR^i, \, \LL^{\bar{\imath}} + \tl \RR^i, \,
\RR^\a + \LL^{\bar\a}, \, \RR^\a - \LL^{\bar \a}) \ . \la{vielbein2}
\ee
Up to permutations, the non-zero components of the metric and $\H$-tensor are
\begin{gather}
\GG_{ij} = \GG_{\bar{\imath}\bar{\jmath}} = \tfrac{\tilde k}{2} \delta_{ij} \ , \qquad \GG_{\a\b} = \tfrac{\tilde k}{2} \delta_{\a\b} \ ,
\\
\H_{\a\b\g} = -\tfrac{\tilde k}{2} f_{\a\b\g} \ , \qquad \H_{\a ij} = -\tfrac{\tilde k}{2} f_{\a ij} \ , \qquad \H_{\a \bar{\imath} \bar{\jmath}} = -\tfrac{\tilde k}{2} f_{\a \bar{\imath} \bar{\jmath}} \ . \la{Hcoset}
\end{gather}
The $\F$ gauge symmetry of the model \rf{214} is manifested in the
vanishing of $\bar{\a}$-components of the metric and the $\H$-tensor.

As discussed in Appendix \ref{B} for the coset \sm{} (see above eq.\rf{b15}),
there are various approaches that can be used to treat the $H$ gauge symmetry.
For example,
we may take $(\hat h, \bar h) \in G\times G$
to be parametrized by the $2\dim G - \dim \F$
physical degrees of freedom and understand $e^{\bar \a} = \RR^\a - \LL^{\bar\a}$ as expanded in the vielbein $(e^i,e^{\bar\imath},e^\a)$.
Alternatively, we can lift the degeneracy of the metric by setting
$\GG_{\bar \a \bar \b} = \e \tfrac{\tilde k}{2} \d_{\a \b}$,
then project out the $\bar{\a}$ directions and finally set the regulator $\e$ to zero.

Using either of these methods to
compute the torsionful spin connection and the corresponding curvature,
the non-zero curvature components are given by
\begin{equation}\begin{split}\la{torsionriem2}
(\tl^{-2}-1){\hat{R}^i}{}_{jkl} =
-\sqrt{\tl^{-2}-1}{\hat{R}^i}{}_{j k \bar{l} }
= {\hat R}^i{}_{j \bar{k} \bar{l} } = f^i{}_{j\a} f^{\a}{}_{kl} \ .
\end{split}\end{equation}
Plugging this
into the RG equation \eqref{rg}, one obtains
\begin{equation}\la{2202}
(\beta^{(1)}_{\m\n} + \beta^{(2)}_{\m\n} )\, dx^\m \otimes dx^\n = \sum^{{\rm dim}\, G - {\rm dim}\, \F}_{i=\bar \imath=1}\Big[ \cg \tl + \frac{\cg}{\tk} \frac{ \tl(\cf - (2\cg - \cf)\tl^2)}{1-\tl^2} \Big] \RR^i \otimes \LL^{\bar{\imath}} \ .
\end{equation}
Here $\cf$ is defined in terms of the index of the representation of $\F$ in which the coset directions transform, as described in Appendix \ref{A}.
In the case that $G$ and $H$ are simple, $\cf$ is proportional to the dual Coxeter number of $H$.

We conclude that only the $\tl$-dependent term in \eqref{214} gets renormalized, i.e.
the 2-loop RG equation \eqref{rg} is solved with $X^\mu=Y_\mu=0$ and
\be \label{02}
\ddt \tk =0 \ , \qquad \qquad \ddt \tl = \frac{\cg \tl}{\tk} \Big[ 1 + \frac{1}{\tk} \frac{ \cf - (2\cg - \cf)\tl^2}{1-\tl^2}\Big] \ .
\ee
The 1-loop term agrees with \cite{Sfetsos:2014jfa,Appadu:2015nfa}.
The level $k$ is RG-invariant, and, as expected, $\tl =\infty,0$ are fixed points.
Expressing $\tk$ and $\tl$ in \eqref{215} in terms of $k$ and the coupling
$\ka$ using \eqref{133} we find
\begin{align}
&\ddt k =0 \ , \la{6d2loopbetacoset}
\qquad\qquad
\ddt \ka = \frac{\cg}{2k} \Big[ (1-\ka^2) + \frac{\cg(1-\ka)^2(1+4\ka-\ka^2) - \cf(1-\ka^4)}{2k\ka} \Big] \ .
\end{align}
The fixed point $\ka=1$ corresponds to $\l=0$ ($\tl = \infty$), that is when the \lam{} reduces to the $G/\F$ gWZW model.
The other fixed point,
which corresponds to the $G/H$ gWZW model with level $-k$,
is corrected and is given by $\ka = -(1+\tfrac{4c_{_G}}{k})$ (cf. \eqref{215}).

As in the group space case \eqref{24} the 2-loop term in the
$\b$-function \rf{02} is, in general, scheme-dependent.
Again, we find that the symmetry under $k \to - k$, $\l \to \l^{-1}$ in \rf{z2sym},
present at the 1-loop order is not there in the 2-loop term of \rf{02}.
However, after introducing the shifted level $\kk = \tk- \cg = k + \cg$ as in \rf{2777}, the 2-loop RG equation for $\tl$ \rf{02} becomes
\be
\ddt \tl = \frac{\cg \tl}{\kk} \Big[ 1 - \frac{1}{\kk} (\cg-\cf)
\frac{1+\tl^2}{1-\tl^2} \Big] \ ,
\ee
which is manifestly invariant under $\kk \to -\kk$, $\tl \to \tl^{-1}$,
a quantum version of the symmetry \rf{z2sym} of the original couplings (cf. \rf{227}).

In the NAD limit \eqref{13} we get from \eqref{6d2loopbetacoset}
\be
\la{555}
\ddt \hh = 2 {\cg} + 4{\cg (\cg - \cf)}{\hh}^{-1} \ ,
\ee
with the 2-loop coefficient being scheme-independent in this 1-coupling limit as in the group space case \eqref{25}.
As verified in Appendix \ref{B}, \rf{555} matches the expression for the
2-loop $\beta$-function of the $G/\F$ symmetric space \sm{}
(reproducing in particular cases the results in \cite{Brezin:1975sq,Hikami:1980hi}).
This demonstrates that the symmetric space \sm{} and its non-abelian dual have the same 2-loop $\b$-function.

\subsection{Low-dimensional examples: \texorpdfstring{$\l$}{lambda}-models for \texorpdfstring{$SU(2)$}{SU(2)} and
\texorpdfstring{$SU(2)/U(1)$}{SU(2)/U(1)}}

Let us first study the simplest case of the model \eqref{6dmodel}, i.e. with $G = SU(2)$.
Parametrizing
\begin{equation}\label{hhbpar}
\tih = \exp(i\phi \sigma_3) \exp(i \theta \sigma_1) \exp(i \psi \sigma_3) \ , \qquad\qquad
\bh = \exp(i\bar \phi \sigma_3) \exp(i \bar \theta \sigma_1) \exp(i \bar \psi \sigma_3) \ ,
\end{equation}
where $\sigma_a$ are the standard Pauli matrices, we find the following 6-dimensional metric and $\BB$-field
($\GG\equiv \GG_{\m\n} dx^\m dx^\n $, \ $\BB\equiv \ha \BB_{\m\n} dx^\m \wedge dx^\n$)
\begin{align}
\GG & = \GG_0 + \GG_1 \ , \qquad \BB = \BB_0 + \BB_1 \ , \no
\\ \no
\GG_0 & = - \tk (d\theta^2 + d\phi^2 + d\psi^2 + 2\cos 2\theta d\phi d\psi +
d\bar\theta^2 + d\bar\phi^2 + d\bar\psi^2 + 2\cos 2\bar\theta d\bar\phi d\bar\psi) \ ,
\\ \no
\BB_0 & = - \tk(\cos2\theta d\phi\wedge d\psi + \cos 2\bar \theta d\bar \phi \wedge d\bar \psi) \ ,
\\ \no
\GG_1 & = - 2\tk\tl \big[\cos 2(\psi+\bar\phi) (d\theta d\bar \theta - \sin 2\theta \sin 2\bar \theta d \phi d\bar \psi) + ( d\psi +\cos2\theta d\phi)(d\bar \phi + \cos 2\bar \theta d \bar \psi)
\\ \no & \qquad \qquad \qquad + \sin 2(\psi+\bar \phi)(\sin 2 \theta d\phi d \bar \theta + \sin 2 \bar \theta d \theta d \bar \psi)\big] \ ,
\\ \no
\BB_1 & = -\tk\tl \big[\cos 2(\psi+\bar\phi) (d\theta \wedge d\bar \theta - \sin 2\theta \sin 2\bar \theta d \phi \wedge d\bar \psi) + ( d\psi +\cos2\theta d\phi)\wedge(d\bar \phi + \cos 2\bar \theta d \bar \psi)
\\ & \qquad \qquad \qquad + \sin 2(\psi+\bar \phi)(\sin 2 \theta d\phi \wedge d \bar \theta + \sin 2 \bar \theta d \theta \wedge d \bar \psi)\big] \ .
\la{su2geom}
\end{align}
One can check explicitly that this metric and $\BB$-field solve the 2-loop RG equation \eqref{rg} with vanishing $X^\mu$ and $Y_\mu$, and the couplings running as in \eqref{01} (with the dual Coxeter number of $G=SU(2)$ given by $\cg = 2$).

Next, let us consider the $\l$-model \eqref{214} for the symmetric space $G/\F = SU(2)/U(1)$.
Using the parametrization for $\tih$ and $\bar{h}$ given in eq.\eqref{hhbpar},
we find that the corresponding
target space background depends on $\psi$ and $\bar\phi$ only
through $\chi = \psi + \bar \phi$, which is a manifestation of the $U(1)$ gauge symmetry.
The resulting 5-d metric and $\BB$-field are
\begin{align} \no
\GG & = \GG_0 + \GG_1 \ , \qquad \BB = \BB_0 + \BB_1 \ ,
\\ \no
\GG_0 & = - \tk (d\theta^2 + d\phi^2 + d\bar\theta^2 + d\bar\psi^2 ) \ , \qquad \BB_0 = 0 \ ,
\\ \no
\GG_1 & = - 2\tk\tl \big[\cos 2\chi (d\theta d\bar \theta - \sin 2\theta \sin 2\bar \theta d \phi d\bar \psi)
+ \sin 2\chi(\sin 2 \theta d\phi d \bar \theta + \sin 2 \bar \theta d \theta d \bar \psi)\big]
\\ \no & \qquad \qquad \qquad + \tk d\chi^2 -2\tk( d\chi +\cos2\theta d\phi)(d \chi + \cos 2\bar \theta d \bar \psi)
\ ,
\\ \no
\BB_1 & = -\tk\tl \big[\cos 2\chi (d\theta \wedge d\bar \theta - \sin 2\theta \sin 2\bar \theta d \phi \wedge d\bar \psi)
+ \sin 2\chi(\sin 2 \theta d\phi \wedge d \bar \theta + \sin 2 \bar \theta d \theta \wedge d \bar \psi)\big]
\\ & \qquad \qquad \qquad -\tk( d\chi +\cos2\theta d\phi)\wedge(d\chi + \cos 2\bar \theta d \bar \psi)\ .
\end{align}
Again one can check explicitly that this metric and $\BB$-field solve the 2-loop RG equation
\eqref{rg} with vanishing $X^\mu$ and $Y_\mu$, and the couplings running as in \eqref{02}
(with the dual Coxeter number for $G=SU(2)$ given by $\cg=2$ and $\cf=0$ for $\F=U(1)$).

\section{Renormalization of \texorpdfstring{$\lambda$}{lambda}-model: standard
configuration space}\label{standard}

In section \ref{extended} we demonstrated the 2-loop renormalizability of the $\lambda$-model \eqref{1} for general groups $G$ and symmetric spaces $G/\F$.
It is then natural to ask what
this implies
for the model on the standard or
physical configuration space, i.e. the \sm{} found by integrating
out $A_\pm$ in \eqref{1}.
Doing so classically gives the following Lagrangian
\begin{equation}\label{lamphys}
\L = k \, \Big( {\PCM}(g) + {\WZ}(g) + \Tr[ J_+\, M^{-1}\, K_- ] \Big) \ ,\quad \qquad M = \operatorname{Ad}_g - I +(1-\lambda^{-1}) P \ .
\end{equation}
Similarly, for the NAD model \eqref{14} we find
\begin{equation}\la{32}
\L = -\tfrac12 \hh\,\Tr\big[ \partial_+ v\, \M^{-1}
\, \partial_- v \big] \ , \quad \qquad \M
= \ad_v + P \ .
\end{equation}
The integration over $A_\pm$ may also give rise to quantum counterterms required to preserve the renormalizability of \eqref{lamphys} at 2 loops \cite{Hoare:2019ark}.
It is natural to expect that, since the term quadratic in $A_\pm$ in the
Lagrangian \eqref{1} has the form $\Tr[ A_+ M A_-]$, these corrections may depend on the matrix $M$ in \rf{lamphys},
but determining their form in general
appears to be
non-trivial.
Here we will focus on the
examples of the \lam{} for the $SU(2)/U(1)$ symmetric space and $SU(2)$ group space.

\subsection{\texorpdfstring{$SU(2)/U(1)$}{SU(2)/U(1)}}

The $\lambda$-model for $SU(2)/U(1)$ is related by analytic continuation to that of $SU(1,1)/U(1)$, which was studied in detail in \cite{Hoare:2019ark}.
Here we briefly summarize certain key points of the discussion there.
Fixing the $U(1)$ gauge symmetry by choosing the following parametrization of the coset element
\begin{equation}
g = \exp(i\alpha\sigma_3)\exp(i \beta \sigma_2) \ , \qquad \cos \a = \sqrt{p^2 + q^2} \ , \qquad \tan \b = \frac{p}{q} \ ,
\end{equation}
the \sm{} \eqref{lamphys} yields the following classical metric
(the $\BB$-field is trivial in 2d target space and $\ka$ is defined in \rf{133})
\begin{equation}\label{g0coset}
\GG_0 = \frac{k}{1-p^2-q^2}(\ka\, dp^2 + \ka^{-1} dq^2) \ .
\end{equation}
The observation in \cite{Hoare:2019ark} was that this metric should be modified by a particular quantum correction
from the determinant \cite{Schwarz:1992te} resulting from integrating over $A_\pm$
\begin{equation}
\delta \GG = -\frac{1}{2}\big(d \log \det M\big)^2 = - \frac{1}{2} \big[d \log (1-p^2-q^2)\big]^2 \ . \la{su2u1counter}
\end{equation}
The 1-loop corrected background $\GG=\GG_0 + \delta \GG$
then solves the 2-loop RG equation \eqref{rg} with
\begin{equation}\begin{split}\label{0oldcoset}
\frac{d}{dt} k & = 0 + \frac{1}{k} \frac{(1-\ka^2)^2}{ \ka^2} \ , \qquad\qquad
\frac{d}{dt} \ka = \frac{1}{k} (1-\ka^2) \ ,
\end{split}\end{equation}
and
\begin{equation}
X^p = -\frac{p}{k\ka}\Big[1-\frac{\ka}{k}+\frac{\ka^{-1}p^2 + \ka q^2}{k(1-p^2-q^2)}\Big]
\ , \quad
X^q = -\frac{q\ka}{k}\Big[1-\frac{1}{k\ka}+\frac{\ka^{-1}p^2 + \ka q^2}{k(1-p^2-q^2)}\Big]
\ , \quad Y_{p,q} = 0.
\end{equation}
In this analysis the symmetry \rf{z2sym}
of the 1-loop RG equation in \rf{0oldcoset}
survives at the 2-loop level.
Indeed, while the leading 1-loop terms in \eqref{0oldcoset} agree with the RG equations
\eqref{6d2loopbetacoset} found from the analysis on the extended configuration space,
they deviate at the 2-loop order.
Since the 2-loop terms in a two-coupling theory are generally scheme-dependent,
we can, in fact, match the two $\b$-function expressions by
redefining the parameters in \eqref{g0coset} as follows
\unskip\foot{\ The most general redefinition achieving this is ($C_1$ and $C_2$ are free constants)
$ k \to k - \tfrac{(1 -\ka)^2}{\ka} + 2 C_1$ and

$\quad \ka \to \ka + \frac{4}{k} \big[(1-\ka)\big(1-(1+\ka)C_2\big) - 2 C_1(1-\ka^2)\operatorname{arctanh}\ka\big]$.
}
\begin{equation}\la{38}
k \to k - \frac{(1 -\ka)^2}{\ka} \ , \qquad \qquad \ka \to \ka + \frac{4(1-\ka)}{k} \ .
\end{equation}
Note that in the $\ka \to 1$ limit the level $k$ remains unmodified, in agreement with this limit corresponding to the $SU(2)/U(1)$ gWZW model.

\subsection{\texorpdfstring{$SU(2)$}{SU(2)}}

Let us now return to the \lam{} for the group $SU(2)$
\cite{Balog:1993es,Sfetsos:2013wia}.
Parametrizing the group element as
\be
g = \exp{\big[ -i \arcsin\a \, \big(\cos{\b} \, \s_2 + \sin{\b}(\cos{\g} \, \s_3 - \sin{\g}\, \s_1)\big)\big]} \ ,
\ee
we obtain from \eqref{lamphys} the following 3d classical \sm{} metric and $\BB$-field \cite{Sfetsos:2013wia}
\unskip\foot{\ The WZ term in $k {\WZ}(g)$ contributes $k(\arcsin\a - \a \sqrt{1-\a^2})$ to the $\BB$-field.}
\begin{equation}\begin{split}\label{b1}
\GG_0 & = k \big[\frac{d\a^2}{\ka(1-\a^2)} + \frac{\kappa \a^2}{\Delta} ( d\beta^2 + \sin^2\beta d\gamma^2)\big] \ , \qquad\qquad
\Delta \equiv \ka^2 + (1-\ka^2)\a^2 \ ,
\\
\BB_0 & = k\big(\arcsin\a - \frac{\ka^2\a\sqrt{1-\a^2}}{\Delta} \big)\sin\beta\, d\beta\wedge d\gamma \ ,
\\
\H_0 & =d\BB_0= \frac{k\a^2}{\sqrt{1-\a^2}\Delta^2}\big[2\ka^2+(1-\ka^2)\Delta\big] \sin\b\, d\a \wedge d\b \wedge d\g \ .
\end{split}\end{equation}
It is possible to ensure the 2-loop renormalizability of the model
by adding to this classical background special
quantum counterterms. The resulting 1-loop ($1/k$) corrected background is
\begin{equation}\begin{split}\label{b2}
\GG & = \GG_0 + \frac{2(1-\ka^2)^2 \a^4}{\ka^2 (1-\a^2)\Delta^2} d\a^2 \ ,
\\
\BB & = \frac{\kk}{k} \BB_0 -2 \big( \arctan\frac{\a}{\ka\sqrt{1-\a^2}}
- \frac{\ka \a \sqrt{1-\a^2}}{\Delta}\big)\sin\beta\, d\beta\wedge d\gamma \ ,
\\
\H & = \frac{\kk}{k} \H_0 - \frac{4\ka\a^2}{\sqrt{1-\a^2}{\Delta^2}} \sin\b\, d\a \wedge d\b \wedge d\g \ ,
\qquad \qquad \kk \equiv k + \frac{4+(1+\ka^2)^2}{4\ka} \ .
\end{split}\end{equation}
This corrected background \eqref{b2} solves the 2-loop RG equation \eqref{rg} with
\begin{equation}\label{0}
\frac{d}{dt}k = 0 + \frac{(1-\ka^2)^3(5+3\ka^2)}{8k\ka^2} \ , \qquad
\qquad \frac{d}{dt} \ka = \frac{(1-\ka^2)^2}{2k} \Big[1 - \frac{(1-\ka^2)^2}{k\ka} \Big] \ ,
\end{equation}
and the only non-zero component of $X^\m$ being ($Y_\m=0$)
\begin{equation}\begin{split}\label{diffv}
X^\a & = \frac{\a(1-\a^2)(1-\ka^2)}{k\Delta}\Big[\ka + \frac{2\a^2\ka^2(1-\ka^2) - (3-2\ka^2+\ka^4)\Delta^2}{k\Delta^2}\Big] \ .
\end{split}\end{equation}
Again, in this analysis the symmetry \rf{z2sym}
of the 1-loop RG equation in \rf{0oldcoset}
survives at the 2-loop level and,
while the 1-loop $\b$-functions for $k$ and $\ka$ match
those found in the extended configuration space approach, i.e. \eqref{24}
with $\cg = 2$, the matching of the 2-loop terms is only achieved after
the following redefinition
\unskip\foot{\ The most general redefinition achieving this is
($C_1$ and $C_2$ are free constants) $k \to k - \frac{4+(1+\ka^2)^2}{4\ka}+2 C_1$
and
$\quad \ka \to \ka - \frac{1}{k} \big[ (1-\ka)^2(1+\ka) (1 + (1+\ka)C_2) + (C_1-1)(1-\ka^2)((1-\ka^2)\operatorname{arctanh}\ka+\ka)\big]$.}
\begin{equation}\begin{split}\label{shiftpcm}
k & \to k - \frac{4+(1+\ka^2)^2}{4\ka}+2 \ , \qquad\qquad
\ka \to \ka - \frac{(1-\ka)^2(1 +\ka)}{k} \ .
\end{split}\end{equation}
As in \eqref{38}, the level $k$ is not modified in the WZW limit $\ka \to 1$.

Note that the coupling $\kk$ defined in \eqref{b2} does not run at 2-loop order,
$\frac{d}{dt} \kk = 0 $.
This is consistent with the fact that $k$ does not run in the extended configuration space
approach and the
particular
shift \eqref{shiftpcm} required to recover \eqref{24} from
\eqref{0}.
This RG invariant $\kk$ is the coefficient of $\arcsin\a$ (present in $\BB_0$ in \rf{b1}) in the 1-loop corrected background \eqref{b2}. Choosing it
to be integer-valued removes the global ambiguities arising from the $\arcsin\a$ term.
Furthermore, given that under a large transformation
\unskip\foot{\ Note that $\arctan\frac{\a}{\ka\sqrt{1-\a^2}} = \operatorname{sign} \ka \, \arcsin\frac{\a}{\sqrt{\Delta}}$ and $\alpha = \frac{\alpha}{\sqrt{\Delta}}$ for $\alpha = 0,\pm1$.}
\begin{equation}
\delta (\arcsin \a) = \operatorname{sign} \ka \ \delta (\arctan\frac{\a}{\ka\sqrt{1-\a^2}})\ ,
\end{equation}
and that the coefficient of $\arctan\frac{\a}{\ka\sqrt{1-\a^2}}$ in \eqref{b2} is integer-valued, this term does not lead to any additional ambiguities.
The quantization of the flux
\be
\frac{1}{4\pi^2} \int \H_0 = k \ , \qquad\qquad \frac{1}{4\pi^2} \int \H = \kk -2 \ ,
\ee
also supports the identification of $\kk$ as integer-valued.

In the case of the $SU(2)/U(1)$ \lam{}
it was possible to write the 1-loop corrections in a simple
way \eqref{su2u1counter} in terms of the matrix $M$ as defined in \eqref{lamphys}.
Let us try to do the same for the 1-loop corrections \eqref{b2} to the $SU(2)$ \lam.
Since the $SU(2)/U(1)$ counterterm \eqref{su2u1counter} took the form $\Delta \L = \ha (\del \log \det M)^2 = \ha (\Tr [M^{-1} \del M ])^2$, we may consider counterterms (with coefficients $c_i$)
built out of the quantity $M^{-1} \del M$.
In addition, we may also include counterterms (with coefficients $d_i$)
proportional to the terms present in the classical Lagrangian \eqref{lamphys} (and its image under parity). As a result, we are led to the following ansatz
\begin{equation}\begin{split}\label{qccc}
\Delta \L &= c_1 (\del_+ \log \det M) ( \del_- \log \det M)
+ c_2 {\PCM}(M) + c_3 {\WZ}(M)
\\ &\quad + d_1 {\PCM}(g) + d_2 {\WZ}(g) + d_3 \Tr[J_+ M^{-1} K_-] + d_4 \Tr[J_- M^{-1} K_+] \ .
\end{split}\end{equation}
We find that this matches the required 1-loop corrected background
\eqref{b2} provided the constants $c_i$ and $d_i$ take the following values
\begin{align}\label{origin}
&c_1 =-\tfrac{3+\ka}{2(1-\ka)} \ , \qquad c_2 = - \tfrac{2(1+\ka)}{1-\ka} \ , \qquad c_3 = 1 \ ,
\\
&d_1 = 0 \ , \quad d_2 = {\rm k} - k - 2 \ ,\quad
\ d_3 = \tfrac{1}{2}\big[{\rm k} - k + (1+\kappa^{-1})\big] \ , \quad d_4 = -\tfrac{1}{2}\big[{\rm k}-k - 3 (1+\kappa^{-1})\big] \ .\no
\end{align}
Combining the quantum counterterms \eqref{qccc} with the classical Lagrangian \eqref{lamphys} allows us to represent the \sm{} corresponding to
the 1-loop corrected geometry \eqref{b2} in the form
\begin{equation}\begin{split}\label{originlag}
\L & = k\, {\PCM}(g) + ({\rm k} -2) {\WZ}(g)
\\ & \qquad
+ \tfrac12 \Tr\Big[ \big({\rm k} + k + (1+\ka^{-1})\big) J_+ M^{-1} K_-
- \big({\rm k}-k - 3 (1+\ka^{-1})\big) J_- M^{-1} K_+ \Big]
\\
&\qquad - \tfrac{3+\ka}{2(1-\ka)} (\del_+ \log \det M)(\del_- \log \det M) - \tfrac{2(1+\ka)}{1-\ka} {\PCM}(M) + {\WZ}(M) \ .
\end{split}\end{equation}
Note that in the WZW limit $\ka \to 1$ when the RG invariant $\kk$ in \eqref{b2} reduces to the usual shift of the level
\begin{equation}
\kk\big|_{\ka = 1} = k+2 \ ,
\end{equation}
the other corrections to the metric and $\BB$-field in \eqref{b2} vanish, so that
the expression in \eqref{originlag} indeed reduces to the standard WZW Lagrangian
$k \big[{\PCM}(g) + {\WZ}(g) \big]$.
\unskip\foot{\ While the coefficients $ \tfrac{3+\ka}{2(1-\ka)}$ and $\tfrac{2(1+\ka)}{1-\ka}$ blow up in the limit $\ka \to 1$, the corresponding $M$-dependent expressions vanish faster.
Note also that $\arcsin \a = \arctan\frac{\a}{\sqrt{1-\a^2}}$.}

\subsubsection{\texorpdfstring{$SU(2) \times U(1)$}{SU(2) x U(1)} invariant limit: abelian T-duality to squashed 3-sphere}

Let us consider the formal limit
\begin{equation}
\alpha \to \sin(\a + i \zeta) \ , \qquad \qquad \zeta \to \infty \ ,
\end{equation}
in which the 1-loop corrected background \eqref{b2} becomes
\unskip\foot{\ We have dropped trivial
(i.e. closed 2-form)
contributions to the $\BB$-field
such that only the $\arcsin\a \to \a + i \z \sim \a$ term in \eqref{b2} gives a relevant contribution in this limit.}
\begin{align}
\label{simpler}
\GG & = \big(\frac{k}{\kappa} + \frac{2}{\ka^2}\big) d\alpha^2 + \frac{k \kappa}{1-\kappa^2} ( d\beta^2 + \sin^2\beta d\gamma^2) \ ,
\\
\BB & = \big[k + \frac{4+(1+\ka^2)^2}{4\ka}\big] \alpha \sin\beta d\beta\wedge d\gamma \ ,
\qquad
\H = \big[k + \frac{4+(1+\ka^2)^2}{4\ka}\big] \sin \beta d\a\wedge d\b\wedge d\g \ .\no
\end{align}
The resulting metric and $\H$-tensor have $SU(2)\times U(1)$ symmetry
(while, as usual, the $\BB$-field cannot be written in a way that is manifestly invariant).
The background \eqref{simpler} solves the 2-loop RG flow equations \eqref{rg} with the parameters running as in \eqref{0} and $X^\m=Y_\m=0$.

Focusing on
the classical part of \eqref{simpler} by taking $k$ large, we may shift
the $\BB$-field by a closed 2-form in order to make translations in $\alpha$ a manifest symmetry
\begin{equation}\begin{split}\label{class}
\GG & = k \Big[ \frac{1}{\kappa} d\alpha^2 + \frac{ \kappa}{1-\kappa^2} ( d\beta^2 + \sin^2\beta d\gamma^2)\Big] \ ,
\qquad\qquad
\BB = k\cos\beta d \alpha\wedge d\gamma \ .
\end{split}\end{equation}
Applying T-duality in the $\alpha$ direction we find the metric of the squashed 3-sphere
(which also has the interpretation of the $\eta$-deformation of the $SU(2)$ PCM, cf. \cite{Kawaguchi:2010jg})
\begin{equation}\la{323}
\tilde \GG = k\Big[ \kappa (d\tilde\alpha - \cos\beta d\gamma)^2 + \frac{ \ka}{1-\ka^2}(d\beta^2 + \sin^2\beta d\gamma^2)\Big] \ .
\end{equation}
To extend this relation to the quantum level,
let us use the 1-loop corrected form of the T-duality transformation given
in \cite{Kaloper:1997ux}
(see also \cite{Parsons:1999ze})
in an alternative scheme related to ours by
\begin{equation}\label{schemechange}
(\GG+\BB)_{\mu\nu} \to (\GG+\BB)_{\mu\nu} + \frac14 \H^2_{\mu\nu} \ .
\end{equation}
Starting with the background \eqref{simpler} and
implementing this scheme change gives
\begin{equation}\begin{split}\label{newscheme}
\GG & = \big(\frac{k}{\kappa} + \frac{2}{\ka^2} + \frac{(1-\ka^2)^2}{2\ka^2}\big) d\alpha^2 + \big(\frac{k \kappa}{1-\kappa^2} + \frac{1-\ka^2}{2}\big) \big( d\beta^2 + \sin^2\beta d\gamma^2\big) \ ,
\\
\BB & = \big(k + \frac{4+(1+\ka^2)^2}{4\ka}\big)\cos\beta d \alpha\wedge d\gamma \ .
\end{split}\end{equation}
This is a special case of a general ansatz
\begin{equation}\begin{split}
\GG & = e^{\varphi} (d\alpha + V_m dx^m)^2 + \GG'_{mn} dx^m dx^n \ , \qquad \GG'_{mn} = \GG_{mn} - e^{\varphi}V_m V_n \ ,
\\ \BB & = W_m dx^m \wedge d\alpha + \tfrac12 \BB_{mn}dx^m \wedge dx^n \ , \qquad \ \ \ x^m = \{\beta,\gamma\} \ ,
\end{split}\end{equation}
where $\varphi$ is a constant, $V_m = 0$ and $\BB_{mn} = 0$.
Then the 1-loop corrected T-duality transformation rules simplify to
($W_{mn} \equiv \partial_m W_n - \partial_n W_m$)
\begin{equation}\label{tdualityrules}
\tilde\varphi = - \varphi + \tfrac{1}{4}e^{-\varphi} W_{mn}W^{mn} \ , \ \ \ \quad
\tilde V_m = W_m \ , \ \ \ \quad
\tilde \GG'_{mn} = \GG'_{mn} \ , \ \ \ \quad
\tilde W_m = 0 \ , \ \ \ \quad
\tilde \BB_{mn} = 0 \ ,
\end{equation}
where the $\frac{1}{4}e^{-\varphi} W_{mn}W^{mn}$ term is the 1-loop (or $\a' \sim \tfrac{1}{k}$) correction \cite{Kaloper:1997ux}.
As a result, the T-dual background is found to be
\begin{equation}\label{2loopdual}
\tilde \GG = \big[k \ka +\tfrac{1}{2} (1+\ka^2)^2\big]\big(d\tilde\alpha - \cos\beta d\gamma\big)^2 + \big[\frac{k \kappa}{1-\kappa^2} + \tfrac{1}{2} (1-\ka^2)\big] \big( d\beta^2 + \sin^2\beta d\gamma^2\big) \ , \quad \ \ \ \td \BB=0 \ ,
\end{equation}
where we have rescaled $\tilde \alpha$.
\unskip\foot{\ As the dual $\BB$-field is vanishing, undoing the scheme change \eqref{schemechange} has no effect.}
This background \eqref{2loopdual} indeed solves the 2-loop RG equations \eqref{rg} with the parameters running as in \eqref{0} and
$X^{\tilde{\alpha},\b,\g} = Y_{\tilde{\a},\b,\g} = 0 .$

On symmetry grounds, the \sm{} for the squashed 3-sphere
\eqref{323} is renormalizable to all loop orders (without the need for counterterms of different form)
\cite{n2} (see also Appendix \ref{B}).
Indeed, this is consistent with \eqref{2loopdual} as the coupling redefinition
\begin{equation}
k = \hat{k} - \frac{3+\hat{\ka}^4}{2\hat{\ka}} \ , \qquad \qquad \ka = \hat{\ka}+\frac{1-\hat{\ka}^2}{\hat{k}} \ ,
\end{equation}
gives simply (cf. \eqref{323})
\begin{equation}\label{2loopdual2}
\tilde \GG = \hat k\Big[ \hat\ka(d\tilde\alpha - \cos\beta d\gamma)^2 + \frac{ \hat\kappa}{1-\hat\kappa^2} ( d\beta^2 + \sin^2\beta d\gamma^2)\Big] \ .
\end{equation}
We thus find another example (in addition to the one discussed in \cite{Hoare:2019ark})
of how the required loop corrections to the T-duality transformation rules
naturally appear
from the deformation under the RG flow of more general integrable models.
Similar higher-loop corrections are expected for non-abelian (and also Poisson-Lie)
duality.

\subsubsection{Non-abelian dual of \texorpdfstring{$SU(2)$}{SU(2)} PCM}

In the case of the $SU(2)$ $\lambda$-model in the coordinates \eqref{b1}, the NAD limit \eqref{13} amounts to
\begin{equation}\label{nadlimit2}
\a \to \ka \alpha\ , \ \ \ \qquad
\ka \to \tfrac{1}{4} \hh k^{-1} \ , \qquad \ \ \ \qquad k \to \infty \ ,
\end{equation}
where $\hh$ and the new coordinate $\a$ are fixed.
In this limit the 1-loop-corrected background \eqref{b2} becomes
\begin{align}
\GG & = \Big[\frac{\hh}{4} + \frac{2\a^4}{(1+\a^2)^2} \Big] d\alpha^2 + \frac{h}{4}\frac{\a^2}{(1+\a^2)} (d\beta^2 + \sin^2\beta d\gamma^2) \ , \no \\
\BB & = \Big[\frac{\hh+5}{4}\frac{\a^3 }{(1+\a^2)} + \frac{2\a}{1+\a^2} - 2\arctan{\a}\Big] \sin \b d\b\wedge d\g \ , \no \\
\H & = \Big[\frac{\hh+5}{4}\frac{\a^2(3+\a^2) }{4(1+\a^2)^2} - \frac{4\a^2}{(1+\a^2)^2}\Big] \sin \b d\a\wedge d\b\wedge d\g \ . \label{f2zero}
\end{align}
It solves the 2-loop RG equations \eqref{rg} with
\begin{align}
&\qquad \qquad \qquad \frac{d}{dt} \hh = 2 + {2}{\hh}^{-1} \ , \la{333}\\
\label{xcorr2}
&X^\a = \frac{2\a}{\hh(1+\a^2)}\Big[(1 -\a^2) - \frac{4(5+5\a^2+3\a^4+\a^6)}{h(1+\a^2)^2}\Big] \ , \qquad X^{\b,\g} = Y_{\a,\b,\g} = 0 \ .
\end{align}
The RG equation \eqref{333} for $\hh$ follows
also by
taking the limit \eqref{nadlimit2} in \eqref{0}. It
matches the 2-loop running of the coupling in the $SU(2)$ PCM (cf. \rf{Agb}).
The limit of the diffeomorphism vector $X^\a$ \eqref{diffv} gives \eqref{xcorr2} after taking into account an extra contribution due to the RG-dependent rescaling of $\a$ in \eqref{nadlimit2}.

We can also take the NAD limit \eqref{13} in the corrected Lagrangian \eqref{originlag},
thus getting
\begin{equation}\begin{split}\label{originlagnad}
\L & = - (\tfrac{1}{2} \hh + \tfrac{9}{4}) \Tr\big[\partial_+ v\, \M^{-1} \partial_- v \big]
- \tfrac{7}{4} \Tr\big[\partial_- v\, \M^{-1} \partial_+ v \big]
\\
&\qquad - \tfrac{3}{2} (\del_+ \log \det \M)(\del_- \log \det \M) - 2 \, {\rm PCM}(\M) + {\rm WZ}(\M) \ .
\end{split}\end{equation}
Setting
\begin{equation}
v = -\tfrac{i}{2} \a \big[\cos{\b} \, \s_2 + \sin{\b}(\cos{\g} \, \s_3 - \sin{\g}\, \s_1)\big] \ ,
\end{equation}
we recover the expected \sm{} with
couplings
given by \eqref{f2zero}.

Taking the further limit of infinite shift of $\a$, i.e.
\begin{equation}\la{limit}
\alpha \to \alpha + \ell \ , \qquad \ \ \ \ell \to \infty \ ,
\end{equation}
in \eqref{f2zero}, we find
\begin{align}
\GG = \tfrac{\hh+8}{4}\,d\alpha^2 + \tfrac{\hh}{4}\big(d\beta^2 + \sin^2\beta d\gamma^2\big) \ ,\qquad \ \ \
\BB = \tfrac{\hh+5}{4} \, \cos \b\, d\a \wedge d\g \label{f2zerolim1}
\ ,
\end{align}
where we have dropped a trivial (closed 2-form)
contribution to the $\BB$-field
in order to make the shift-symmetry of $\a$ manifest.
The resulting background is thus $\mathbb{R} \times S^2$ supported by a constant $\H$-tensor.
Changing scheme as in \eqref{schemechange},
T-dualising in $\a$ using the rules in eq.\eqref{tdualityrules} and rescaling $\tilde \a$
we find
\unskip\foot{\ Note that, since we now have a 1-coupling theory, the 2-loop $\beta$-function \eqref{333} is scheme-independent and the shift of $\hh$ here is not in contradiction with the results above. Indeed, sending $\hh \to \hh + 2$ leaves \eqref{333} invariant to 2 loops.}
\begin{equation}\begin{split}
\tilde \GG & = \tfrac{\hh+2}{4}\big(d\b^2 + d\tilde{\a}^2 + d\g^2 - 2\cos\b\, d\tilde\a d \g\big)
= (\hh+2) \big( d \theta ^2+ \sin^2 \theta\, d \psi^2 + \cos^2 \theta \, d \chi^2 \big) \ ,
\end{split}\end{equation}
i.e. the $SU(2)$ PCM with the $S^3$ radius-squared equal to $\hh+2$.
\unskip\foot{\ Note that in the case when $S^3$ is interpreted as a coset
$SO(4)/SO(3)$ the coupling of the symmetric space \sm{} is given by
$\hh= 2 R^2$ (cf. footnote \ref{f34}).}
Here $\tilde \a = \psi + \chi$, $\g = \psi - \chi$ and $\b = 2\theta$.
Let us also note that in the limit
$\a \to \ve \a , \, \hh \to \ve^{-2}\hh , \ \ve \to 0 ,$
the background \eqref{f2zero} becomes flat with vanishing $H$.

\section{Concluding remarks}\label{conclusions}

As we have seen above, formulating the \lam{} on extended
($G\times G\times G$)
configuration space
``linearizes'' the RG flow, i.e. makes its renormalizability manifest without the need
for extra local counterterms apart from running of the coupling $\l$.
The same is true in the limit \eqref{13} that gives the
interpolating model for non-abelian duality.
\unskip\foot{\ A similar approach may also be useful
for clarifying the higher-loop deformation in abelian T-duality.
In this case the model on the ``tripled'' configuration space can be found
from the interpolating model for abelian T-duality.
For example, consider the metric $\GG = dy^2 + a(y) dx^2$ and its classical dual
$\td\GG = dy^2 + a^{-1} (y) d\td x^2$
with the interpolating model given by
$L= (\del_r y)^2 + a(y) (A_r)^2 + \td x \eps^{rs} F_{rs} $
such that $y$ is a spectator coordinate.
If we integrate out $\td x$ to give $A_r = \del_r x$ we recover the
original model for $x$.
If we integrate out $A_r$ we find the T-dual model for $\td x$.
If instead we set $A_1=\del_1 x $ and integrate out $A_0$ we get the
``doubled'' model of \cite{Tseytlin:1990va} for $x$ and $\td x$ (equivalent
to the ``axial'' gauge choice in the Appendix of \cite{Rocek:1997hi}).
The ``tripled'' model for ($x, \td x, \bar x$) is obtained by setting $A_r= \del_r x + \eps_{rs}\del^s \bar x$:
$\L= (\del_r y)^2 + a(y) [ (\del_r x)^2 - (\del_r \bar x)^2 + 2 \eps^{rs} \del_r x \del_s \bar x ] + \del_r \td x \del^r \bar x $.
This may be interpreted as a \sm{} on 4-dimensional target space with pp-wave metric and $\BB$-field.
}
Using this relation we demonstrated that the PCM and symmetric space \sm{} have the same 2-loop $\beta$-functions as their
non-abelian duals, thereby extending their quantum equivalence to the 2-loop level.

One open problem is how to interpret the local counterterms
required for 2-loop renormalizability of the \lam{} defined on the standard configuration space \eqref{lamphys}
starting from the manifestly renormalizable theory on the extended configuration space \eqref{22}.
In the simplest example of $SU(2)/U(1)$ model the origin of the counterterm \eqref{su2u1counter}
can be traced to the determinant resulting from integrating out the 2d gauge field \cite{Hoare:2019ark}.
However, in the $SU(2)$ model with 3d target space the derivation and structure of
the rather intricate counterterms in \eqref{originlag} and \eqref{originlagnad} are not immediately clear.

Another interesting question is to understand how integrability implies renormalizability
and if renormalizable $\s$-models should always be integrable.
\unskip\foot{\ Here we consider only $\s$-models without potential terms.
Adding potentials one can certainly arrange to have renormalizability in perturbation theory without
having integrability.
So, in general, integrability may imply renormalizability but not vice versa.}
Whether this relationship should be with classical or quantum integrability is also of interest.
Indeed, there are well-known cases in which the classical integrability is
anomalous, e.g. the bosonic $\mathbb{C}\mathbf{P}^N$ model \cite{Abdalla:1980jt} (see also \cite{ey1}).
However, this does not appear to be reflected in the 2-loop renormalizability.
For the $\mathbb{C}\mathbf{P}^N$ model it has been conjectured that quantum integrability
can be restored by including an additional free field \cite{cpn1} in the classical limit (related models have also appeared in \cite{bbr}).
The precise way in which this occurs and how it can be understood in the ``tripled'' configuration space remain to be understood.
An alternative is to consider the supersymmetric $\mathbb{C}\mathbf{P}^N$ \sm{} in which there is no anomaly \cite{Gomes:1982qh}.
It could also prove insightful to redo the analysis in this paper for such models.

A potentially useful application of our results is to the $\eta$-model of \cite{klimcik,dmv}, the one-loop renormalisability of which was studied in \cite{Squellari:2014jfa} (see also \cite{oneloopeta}). 
Up to analytic continuation, the $\eta$-model and \lam{} are related by limits and T-duality \cite{Hoare:2015gda,hs}
or by Poisson-Lie duality \cite{Hoare:2015gda,pld,hs}.
These connections may be used to investigate both the renormalizability of the $\eta$-model at higher loops and corrections to non-abelian and Poisson-Lie duality. 
In our analysis of the \lam{} we have computed the 2-loop $\beta$-function of the models \eqref{6dmodel}
and \eqref{214}. It would be interesting to extend this to the more general ``doubly $\lambda$-deformed'' models
constructed in \cite{Georgiou:2016urf,Georgiou:2017aei,Georgiou:2017oly,Georgiou:2017jfi}.

Another obvious generalization is to $\s$-models on supergroups and
supercosets (see, e.g.,
\cite{Kagan:2005wt,Babichenko:2006uc,Zarembo:2010sg,Appadu:2015nfa}) where a
generalization of our 2-loop $\beta$-function expressions would, e.g., check
the 2-loop finiteness of the model of \cite{Hollowood:2014qma}.

\section*{Acknowledgments}

We would like to thank R. Borsato, S. Demulder, F. Seibold, K. Sfetsos, D. Thompson and L. Wulff for useful discussions.
We also thank K. Sfetsos for drawing our attention to ref. \cite{Georgiou:2017oly}.
BH was supported by the Swiss National Science Foundation through the NCCR SwissMAP.
NL was supported by the EPSRC grant EP/N509486/1.
AAT was supported by the STFC grant ST/P000762/1.

\bigskip

\appendix

\section{Notation and conventions} \label{A}
\def\theequation{A.\arabic{equation}}
\setcounter{equation}{0}

Our conventions for the non-zero
components of the 2d Minkowski metric $\eta_{rs}$ and the Levi-Civita symbol $\epsilon^{rs}$
are, respectively, $\eta_{00} = -\eta_{11} = -1$ and $\epsilon^{01} = -\epsilon^{10} = 1$.
The 2d light-cone coordinates are defined as $\sigma^\pm = \ha (\sigma^0 \pm \sigma^1)$
so that $\del_\pm = \del_0 \pm \del_1$.
For $n$-forms we define components in the standard way: $\H_{(n)} = \tfrac{1}{n!}\H_{i_1\cdots i_n} dx^{i_1} \wedge \cdots \wedge dx^{i_n}$.

For an irreducible finite-dimensional representation of the compact simple Lie group $G$ we normalize the generators $\{T^a\}$ and the invariant bilinear form such that
\unskip\foot{\ This means that we have hermitian generators and imaginary structure constants.}
\be\la{a1}
[ T^a, T^b ] = f^{ab}{}_{c} T^c \ , \qquad
\Tr (T^a T^b) = \d^{ab} \ , \qquad f^{ab}{}_{c} f_{abd} = - 2\cg\d_{cd} \ , \qquad \cg\equiv c_2(G)\ , \ee
where $c_2(G)$ is the dual Coxeter number of the group $G$
and indices are raised and lowered with $\d^{ab}$ and its inverse.
This implies that $\Tr$ is related to the usual matrix trace, $\tr$, by
\begin{equation}\la{trnorm}
\Tr = \tfrac{1}{2\chi_{_{G,R}}}\tr \ ,
\end{equation}
where $\chi_{_{G,R}}$ is the index of the representation.
\unskip\foot{\ \la{con} We use a
somewhat unconventional normalization of the generators and thus the structure constants (by a factor of $\sqrt 2$)
compared to the standard relations
$
\tr (T'^a T'^b) = \chi_{_{G,R}} \delta^{ab} $ and $f'^{ab}{}_{c} f'_{abd} = - \cg\d_{cd}$.
Nevertheless, our normalizations are consistent with the standard values for the indices of representations.
For the fundamental representation we have $\chi_{_{SU(N),\textrm{fund}}} = \chi_{_{Sp(N),\textrm{fund}}} = \tfrac12$ and (for $N\geq5$) $\chi_{_{SO(N),\textrm{fund}}} = 1$,
while for the adjoint representation the index is equal to the dual Coxeter number: $\chi_{_{SU(N),\textrm{adj}}} = c_2(SU(N)) = N$, $\chi_{_{Sp(N),\textrm{adj}}} = c_2(Sp(N)) = N+1$ and (for $N \geq 5$) $\chi_{_{SO(N),\textrm{adj}}} = c_2(SO(N)) = N-2$.}
It then follows that
\be \la{a2}
f_{a}{}^{de} f_{be}{}^{f} f_{cdf} = \cg f_{abc}\ .
\ee

In the coset case, $G/\F$ is assumed to be a compact irreducible symmetric space.
Introducing the orthogonal splitting $\{T^a\}= \{T^\alpha, T^i\}$, where $\{T^\a\}$ are the generators of $\F$, we have $\Tr (T^\a T^i) =0 $ and the
non-zero commutation relations are given by
\be \la{a3}
[T^\a, T^\b]= f^{\a\b}{}_{\g} T^\g \ , \qquad
[T^\a, T^i]= f^{\a i}{}_{j} T^j \ , \qquad
[T^i, T^j]= f^{ij}{}_{\a} T^\a \ .
\ee
In the computation of the 2-loop RG flow we make use of the identities
\begin{equation}\begin{split} \la{a10}
f^{\a i}{}_{j}f^{\b j}{}_{i} & = 2\chi_{_{H,G/H}} \delta^{\a\b} \ ,
\qquad
f^{\a k}{}_i f_{\a k j} = -c_{_G} \d_{ij}\ ,
\\
f^{\a ml} f^{\b k}{}_l f_{\a ki} f_{\b mj}
& = -\tfrac12 f^{\a m}{}_{l}f^{\b l}{}_{m} f_{\a}{}^{k}{}_i f_{\b kj} = \cg (\cg -\cf) \delta_{ij} \ , \qquad
\cf \equiv \cg - \chi_{_{H,G/H}} \ ,
\end{split}\end{equation}
where $\chi_{_{H,G/H}}$ is the index of the representation of $\Lie(\F)$ in which the coset directions transform, i.e. of the matrices $(f^\a)^i{}_j = f^{\a i}{}_j$.
When both $G$ and $\F$ are simple we have that $\cf \equiv \tfrac{\chi_{_{G,R}} }{\chi_{_{\F,R}}}c_2(\F)$
where $c_2(\F) $ is the dual Coxeter number of the group $\F$ and $\chi_{_{\F,R}}$ is the
index of the representation $\{T^\a\}$ of $\Lie(H)$.
For more general subgroups $\F$ the constant $\cf$ takes a more complicated form.
In the case of type II symmetric spaces, i.e. $\frac{\F\times \F}{\F}$, we have $\cg = c_2(\F)$ and $\cf =\ha {c_2(\F)}$.
Finally, the expressions for the dual non-compact irreducible Riemannian symmetric space can be found by the formal substitution $k \to - k$.
\unskip\foot{\ In the classification of irreducible Riemannian symmetric spaces, excluding the special case of flat space, every compact space has a corresponding non-compact space, often referred to as a duality.
The simplest example of this is the sphere and the hyperboloid.
The non-compact irreducible Riemannian symmetric spaces take the form $G/\F$ with $\F$ the maximal compact subgroup of $G$.
Therefore, the coset directions are all non-compact and for a positive-definite signature of the metric we replace $k\to-k$ compared to the compact case.}

\section{2-loop \texorpdfstring{$\beta$}{beta}-function of squashed PCM and \texorpdfstring{$G/\F$}{G/H} coset \texorpdfstring{$\s$-models}{sigma-models}} \label{B}
\def\theequation{B.\arabic{equation}}
\setcounter{equation}{0}

Let $G$ and $\F \subset G$ be compact simple Lie groups and $G/\F$ be a compact irreducible symmetric space.
Below we shall consider the renormalisation of the ``squashed'' PCM model
with action $\S = \frac{1}{ 4 \pi} \int d^2 \s \, \L $ where
\unskip\foot{\ As in Appendix A (cf. \rf{a3})
we denote the $\F$ and $G/\F$ algebra indices by $\a$ and $i$ respectively, i.e. the $G$ algebra index is $a=\{\a,i\}$.
The overall minus sign is due to the conventions explained in Appendix \ref{A},
in particular the choice to use hermitian generators $T^a$ satisfying \rf{a1}.
The coupling $\hh$ is related to the conventional PCM coupling $\rm g$ by
$\hh = \tfrac{2}{{\rm g}^2}$.
Indices in \eqref{bb1} and \eqref{Amet} are contracted with $\delta_{ij}$ and $\delta_{\a\b}$.}
\begin{align}
& \L =- \ha \hh \big( J_+^i J_-^i + \e J_+^\a J_-^\a \big) = -\ha \big(\hh J_+^i J_-^i+
\hhh J_+^\a J_-^\a \big)
\ , \la{bb1}\\
&\qquad J^a_\pm = (J^\a_\pm , J^i_\pm)= \Tr ( T^a g^{-1} \del_\pm g) \ ,\qquad \qquad \hhh \equiv \hh\, \e \ . \la{bb101}
\end{align}
It interpolates between the PCM on the group $G$ ($\e =1$) and the $G/\F$ symmetric space \sm{} ($\e \to 0$).
Also, in the limit
\be
\hh \to \infty\ , \qquad \e\to 0 \ , \qquad \qquad \hhh={\rm fixed} \ , \la{b12}
\ee
when the coset part decouples, the model \rf{bb1} reduces to the PCM on the group $\F$ with the coupling $\hhh$.

For $\e\neq 0$, the action \rf{bb1} has global $G \times H$ symmetry: $ g \to ugv$, $u \in G$, $v\in H$. For the symmetric space \sm{}
case of $\e=0$, the global $H$ symmetry is enhanced to a gauge symmetry. Due to these symmetries
\unskip\foot{\ This follows from the assumption that $\F$ is simple and $G/\F$ is an irreducible symmetric space.
Therefore both $\{T^\a\}$ and $\{T^i\}$ transform in irreducible representations of $\Lie(\F)$ and the two terms in \rf{bb1} are the only ones that respect the global $G \times \F$ symmetry of the model.
In principle, this argument applies to any coset space $G/\F$ for which this irreducibility of representations holds.
At the $\e = 0$ point
we recover the symmetric space or coset \sm{} and
the $\F$ gauge symmetry implies that just the irreducibility of $\{T^i\}$ is sufficient for renormalizability.
An additional subtlety can occur when the irreducible representations are reducible over $\mathbb{C}$.
Then it may be possible to construct new terms respecting the symmetries invoked above (cf. footnote \ref{caveat}). \la{caveat2}
We expect any such new terms to be excluded by additional symmetries such as parity.}
the model \rf{bb1}
is renormalizable with only the two couplings $\hh$ and $\e$ (or $\hhh$) running.
This will be explicitly verified below in the 2-loop approximation (expanding in large $\hh$
for fixed $\e$).

We define the target space vielbein
$E^a \equiv E^a_\mu dx^\mu= (J^i, J^\a)$
where $J^a$ are 1-forms corresponding to the currents in \rf{bb1} (cf. \eqref{vielbein},\eqref{vielbein2}), so that the corresponding
metric of the \sm{} in \rf{15} takes the form
\be
\GG_{\m\n} dx^\m dx^\n = - \ha \hh \big( J^i J^i + \e J^\a J^\a \big) \ .
\la{Amet}
\ee
The spin connection ${\w^a}_b$ is found to have the following components
\be
{\w^\a}_\b = -\ha f^{\a}{}_{\b\g} J^\g \ , \qquad {\w^\a}_i = -\ha f^{\a}{}_{i j} J^j \ , \qquad {\w^i}_j = -(1-\ha \e)f^{i}{}_{j\a}J^\a \ ,
\ee
and the non-zero components of the
corresponding Riemann tensor are given by (up to permutations and symmetries)
\begin{align}
&{R^\a}_{\b\d\e} = \tfrac{1}{4} f^{\a}{}_{\b c}f^c_{\ \d\e} \ , && {R^\a}_{i \g k} = \tfrac{1}{4} ( f^{i}{}_{ k c}f^c{}_{\a \g} - \e f^{\a}{}_{kc}f^c{}_{i \g}) \ , \no
\\
&{R^i}_{j\b\g} = \tfrac{1}{2}\e(1-\tfrac{1}{2}\e)f^i{}_{jc}f^c{}_{\b\g} \ , && {R^i}_{jkl} = (1-\tfrac{3}{4}\e) f^i{}_{jc}f^c{}_{kl}\ . \la{Ariem1}
\end{align}
In the PCM case of $\e=1$, the curvature \eqref{Ariem1} reduces to the
standard group space expression
\begin{equation}\la{b5}
{R^a}_{bde} = \tfrac{1}{4} f^a{}_{bc} f^c{}_{d e} \ .
\end{equation}
For $\BB_{\m\n}=0$ the 2-loop RG equation \eqref{rg}
becomes the familiar one \cite{Friedan:1980jm}
\unskip\foot{\ Here we ignore the diffeomorphism term which is not allowed by the global $G \times \F$ symmetry.}
\be \la{b7} \frac{d \GG_{\m\n}}{dt}
= R_{\m\n} + \tfrac12 R_\m ^{\ \lambda \rho \sigma} R_{\nu \lambda \rho \sigma}
\ . \ee
Substituting in \rf{Ariem1}, one
finds that the model \eqref{bb1} is 2-loop renormalizable with $\hh$ and $\e$ running according to
\unskip\foot{\ Note that these equations depend on
$\hh$, $\cg$ and $\cf$ only through the ratios $ \cg \hh^{-1}$ and $\tfrac{\cf}{\cg}$.}
\begin{align}
&\ddt \hh = (2 - \e ) \cg + \ha \cg \hh^{-1} \big[
\cg(8-12\e+5\e^2) - \cff (1 - \e) ( 5 - 4 \e)
\big]\ ,
\la{b19} \\
&\ddt \e =- {\hh}^{-1} (1-\e) \Big( 2 \cg \e - (1 + \e) \cff \no \\
&\qquad \qquad \qquad\qquad + \ha \hh^{-1} \big[ 4 \cg^2 \e (2 - \e) - \cg\cff \e (11-4 \e ) - \cff^2 (\e^{-1} + 1 - 5 \e + \e^2 ) \big] \Big) . \la{b20}
\end{align}
Thus $\e=1$ is a fixed line of \eqref{b20}, on which \rf{b19} reduces to
the 2-loop $\beta$-function of the PCM on the group $G$
\begin{align}
\ddt \hh = \cg + \ha \cg^2 \hh^{-1} \ . \la{Agb}
\end{align}
The same expression was found from the NAD limit of the \lam{} in \eqref{25}.

In the special case of the squashed 3-sphere or squashed $SU(2)$ PCM
(with $H= U(1)$, i.e. $\cg=2, \ \cff=0$) the expressions in
\rf{b19},\rf{b20} agree with the 2-loop $\b$-functions found in \cite{n2}: the couplings $\lambda$ and $g$ used there
are related to ours by $\cg \hh^{-1} = \tfrac{\l}{4\pi},\ \e= 1+g$.

Written in terms of the couplings $\hh$ and $\hhh$ in \rf{bb1} the equations \rf{b19},\rf{b20}
take the form
\begin{align}
&\ddt \hh = 2 \cg +\ha \cg ( 8 \cg - 5 \cf - 2 \hhh) \hh^{-1}
- \tfrac{3}{2} \cg ( 4 \cg - 3 \cf) \hhh \, \hh^{-2}
+ \ha \cg ( 5 \cg - 4 \cf) \hhh^2\, \hh^{-3} , \la{b100} \\
&\ddt \hhh = \cff + \ha \cff^2 \hhh^{-1}
+ (\cg-\cf) \hhh\, \hh^{-2} \big[
3 \cff + \hhh - 3 \cff \hhh \hh^{-1} + \ha (\cg + \cff)\hhh^2 \hh^{-2}
\big] \ . \la{b101}
\end{align}
These two equations become the same and equal to \rf{Agb} at the PCM fixed point $\hh=\hhh$ ($\e=1$).
In the limit \rf{b12} when $\hh^{-1} \to 0$ we get from \rf{b101}
the correct 2-loop RG equation for the coupling $\hhh$
of the PCM on the group $H$ (cf. \rf{Agb})
\unskip\foot{\ Expressed in terms of the physical coupling $\hh^{-1}$
eq.\rf{b100} is $ \ddt \hh^{-1} = - 2 \cg \hh^{-2} + \ldots$ so $\hh^{-1} =0$ is (trivially) a fixed line.}
\be\la{b.11}
\ddt \hhh= \cff + \ha \cff^2 \hhh^{-1} \ .
\ee

Let us now consider the coset space limit $\ep=0$ (or $\hhh=0$).
In the abelian $\F$ case when $\cf=0$, we have $\e=0$ solving \rf{b20} while
\rf{b19} reduces to
$\ddt \hh = 2 \cg + 4 \cg^2\hh^{-1} $, which agrees with the symmetric space \sm{} $\b$-function in \rf{555},\rf{b555}.
However, for non-abelian $\F$ with $\cf\not=0$,
the $\e\to 0$ limit of \rf{b20} is singular.
The limit of \rf{b19},
while giving the correct 1-loop part of the $\b$-function for the $G/\F$ symmetric space \sm{}
(see, e.g., \cite{Appadu:2015nfa}),
fails to do so at the 2-loop order.
Indeed this limit is subtle and should be treated separately due to the $\F$ gauge symmetry that arises in the model \rf{bb1} at the point $\e=0$.

This gauge symmetry is reflected in the degeneracy of the metric \eqref{Amet} in the $\e\to 0$ limit.
One option is to set $\e=0$ and then use the standard gauge fixing procedure. For example, we may fix
an analog of ``axial'' gauge
$N^r J^\a_r = f^\a(\s)$ where $N^r$ is a constant 2d vector. Then averaging over $f$ with exponential weight $ \sim u f^\a f^\a$
will give an extra gauge-fixing term $u ( N^r J^\a_r )^2$ in the coset \sm{} action.
The resulting on-shell effective action and thus the on-shell UV divergences
should not depend on the value of the gauge-fixing parameter $u$ or the
choice of $N^r$.
As this procedure is somewhat cumbersome, we may try to use a short-cut.

Indeed, observing that averaging over $N^r$ should effectively restore 2d Lorentz invariance in gauge-invariant expressions
we may simply add $u (J^\a_r) ^2$ or, equivalently, go back to \rf{bb1} with $u=\e$.
This may be viewed as using
$\e \ll 1 $ as a regulator, breaking the gauge invariance and lifting the degeneracy.
Then after computing the Riemann tensor,
we will need to project out the components in the degenerate $\F$ (or $\alpha$) directions
and finally take the $\e \to0$ limit and compute the $\b$-function.
\unskip\foot{\ In a systematic gauge-fixing the analog of $\e$ or the gauge-fixing parameter $u$
should automatically disappear from the on-shell divergences. Note also that this procedure
is effectively equivalent to fixing a ``transverse'' gauge in
which the $\F$-components of the quantum fields are set to zero so the curvature tensor coefficients in the \sm{} vertices
are contracted with the propagators containing projectors to $G/\F$. In addition, the classical fields (in the background field method for computing divergences)
have only $G/\F$ components due to the classical gauge invariance in the $\e=0$ limit.}
Projecting out the $\a$ components of \eqref{Ariem1} and setting $\e=0$ gives the standard expression for the
symmetric space Riemann tensor (see, e.g., \cite{Castellani:1999fz})
\be
{R^i}_{jkl} = f^i{}_{j\a} f^{\a}{}_{ k l} \ . \la{b15}
\ee
An alternative approach (equivalent to explicitly solving the gauge condition rather than adding it
to the action to lift the degeneracy)
would be to take $g$ to be parametrized by the $\dim G - \dim \F$ physical degrees of freedom.
The particular parametrization is not important, but one could
take, e.g., $g = \exp
\big(i v_i T^i + i v_\a(v_i) T^\a\big)$.
Then we may expand $J^\a$ in the vielbein $E^i = J^i$, i.e. $J^\a = F^\a_i(g) \, E^i$.
Computing the spin correction and the corresponding curvature, the latter does not depend on $F^\a_i(g)$, as expected by gauge invariance, and agrees with \eqref{b15}.

Plugging \eqref{b15} into the RG equation \eqref{b7} we then obtain the 2-loop $\beta$-function
for the symmetric space \sm\
\be
\ddt \hh = 2 \cg + 4 \cg( \cg-\cff) \hh^{-1} \ . \la{b555}
\ee
This expression agrees with previous results found for particular cosets in
\cite{Brezin:1975sq,Hikami:1980hi,Friedan:1980jm}.
\unskip\foot{\ \la{f34} For example,
in the case of the sphere $S^{N-1} = SO(N)/SO(N-1)$,
using the fundamental representation of $SO(N)$ and following the notation in
Appendix \ref{A} we have $\cg = N-2$, $\cf = N-3$, $\chi_{_{G,\textrm{fund}}} = \chi_{_{H,\textrm{fund}}} = 1$.
Normalizing the Lagrangian as in \rf{14},\eqref{15} we find that $\hh = 2 R^2$ where $R$ is the radius of the sphere.
Therefore, from \eqref{b555} we find the standard 2-loop RG equation for $S^{N-1}$: \
$\ddt R^2 = (N-2) + (N-2){R}^{-2}$.}
It also matches the result found from the NAD limit of the \lam{} in \eqref{555}.
In contrast to \rf{b19},\rf{b20}, the expression for the $\b$-function for $\hh$ in \eqref{b555}
is valid for any compact irreducible symmetric space (cf. footnote \ref{caveat2}) with $c_{_G}$ and $c_{_H}$ defined in Appendix \ref{A}, i.e. $G$ and $H$ need not be simple.\foot{\ Let
us note for completeness that renormalizability of \sms{} on some homogeneous cosets (not necessarily symmetric spaces)
was discussed, e.g., in \cite{bonneau}.}

Let us note that since \rf{bb1} is a two-coupling theory, the 2-loop (and higher) terms in the $\b$-functions \rf{b19},\rf{b20}
are not, in general, invariant under scheme changes or redefinitions of the couplings $\hh$ and $\e$.
However, they still contain some invariant information as the limits $\e=1$ (PCM on $G$), \rf{b12} (PCM on $\F$) and $\e=0$ ($G/\F$ coset space)
lead to one-coupling models whose 2-loop $\b$-functions are invariant under coupling redefinitions.

We finish with a curious observation that
the 2-loop $\b$-functions \rf{b19},\rf{b20} vanish if
\be\la{b170} \e=2 \ , \qquad \qquad \cg =\tfrac{3}{ 4} \cf \ . \ee
Indeed, if there are such $G$ and $H$ that the relation $\cg = \tfrac{3}{4} \cf$ can be satisfied, then
the expressions in \rf{b19},\rf{b20} simplify to
\begin{align}
&\ddt \hh = (2 - \e ) \cg\big[ 1 + \tfrac{1}{6} \cg \hh^{-1} ( 2 +\e ) \big]
\ ,
\la{b197} \\
&\ddt \e = \tfrac{2}{3} {\hh}^{-1} (2-\e) (1-\e) \cg \big[ 1 + \tfrac{1}{3} \cg \hh^{-1} ( 2 \e^{-1} + 3 - \e ) \big] \ , \la{b207}
\end{align}
and thus $\e=2$ is a 2-loop fixed point.
This suggests that the corresponding squashed PCM with $\e=2$
is an exact CFT for any value of $\hh$.
Therefore, it is a particularly interesting representation theory question as to whether there are solutions to the condition $\cg = \tfrac{3}{4} \cf$.
For this it may be necessary to consider 
supergroups.



\end{document}